\documentclass[journal]{IEEEtran}
\usepackage[utf8]{inputenc}
\usepackage[T1]{fontenc}
\usepackage{graphicx}
\usepackage{booktabs}
\usepackage{array}
\usepackage{calc}
\usepackage{textcomp}
\usepackage{amsmath,amssymb}
\usepackage{url}
\usepackage[hidelinks]{hyperref}
\providecommand{\textquotesingle}{\textquoteright}
\providecommand{\real}[1]{#1}

\begin{document}
\title{What Went Wrong with Data Lakes?\\ A 15-Year Reality Check from the Field}
\author{\textbf{Youssef~GAHI}\\[3pt]
{\normalsize Laboratory of Engineering Sciences (Laboratoire des Sciences de l'Ing\'enieur)}\\[1pt]
{\normalsize National School of Applied Sciences (ENSA), Ibn Tofail University, Kenitra, Morocco}\\[1pt]
{\normalsize \textit{youssef.gahi@uit.ac.ma}}}
\markboth{Preprint, 2026}{Gahi: What Went Wrong with Data Lakes? A 15-Year Reality Check from the Field}
\maketitle

\begin{abstract}
James Dixon introduced the Data Lake in 2010. The pitch was simple: store data raw, postpone schema design, spend less on up-front transformation. It promised flexibility and easier access to analytics. Fifteen years on, that promise has mostly gone unmet. Survey after survey reports high failure rates, whether the object is a big data program, a Data Lake, or a data science effort meant to reach production. This paper asks why. We read 64 sources spanning academic work, analyst reports, and practitioner accounts, and seven recurring anti-patterns kept reappearing. We call them the Seven Deadly Sins of Data Lakes. We then offer an explanation for them: Governance Debt, the compounding cost of governance decisions that organizations keep putting off. A second pattern surfaced on its own. When governance gets hard, organizations drift back toward structured, warehouse-style approaches. We name that pull governance gravity. The term Data Swamp, meanwhile, is used loosely in the literature, so we give it a working definition with measurable indicators and add a qualitative rubric, the Governance Debt Assessment Model, for catching decay early. The root causes are organizational far more than technical. We also checked whether the newer paradigms, Data Lakehouse and Data Mesh, absorbed the lesson. The technology has advanced. The organizational record has barely moved. For practitioners we provide two tools, a Reality Check Framework and a Stage-Based Intervention Matrix. The paper rests on more than the analyst literature. It draws on a primary source: a catalogue of close to five hundred field reality checks I recorded over fifteen years of building and rescuing enterprise Data Lakes, mostly in financial services and telecommunications across Morocco and West Africa. Assembled independently of that literature, the catalogue lands on the same anti-patterns, brings out two failure dimensions the literature under-reports, operational debt and engineering-discipline debt, and reads the problem from an emerging-market vantage.
\end{abstract}

\begin{IEEEkeywords}
Data lake, data governance, data swamp, big data, data lakehouse, data mesh, governance debt, emerging markets, critical retrospective, practitioner study.
\end{IEEEkeywords}

\section{INTRODUCTION}

"If you think of a datamart as a store of bottled water, cleansed and
packaged and structured for easy consumption, the Data Lake is a large
body of water in a more natural state. The contents of the Data Lake
stream in from a source to fill the lake, and various users of the lake
can come to examine, dive in, or take samples" {[}1{]}. With these
words, James Dixon, then CTO of Pentaho, introduced the Data Lake
concept in October 2010. This elegant metaphor captured the imagination
of the industry and promised a fundamental shift in how organizations
would manage and derive value from their data.

Fifteen years on, the Data Lake has achieved widespread adoption in
enterprise data architectures. The evidence, though, suggests the
paradigm has largely failed to deliver on its promises, and the headline
numbers are hard to ignore. Gartner estimated that 60\% to 85\% of big
data projects fail {[}2{]}, and warned as early as 2014 that, without
governance, data lakes would turn into ``data swamps'' {[}3{]}. The
NewVantage Partners 2020 survey found that although 98.8\% of Fortune
1000 companies were investing in data initiatives, only 37.8\% reported
having built a data-driven organization {[}4{]}. Beyond Gartner and
NewVantage, Table I aggregates complementary indicators across different
data initiatives and use cases, Data Lakes, big data programs, and data
science productionization, all pointing to the same gap between ambition
and outcome. The picture is consistent: across initiatives, results fall
short of the stated ambitions. We examine the lake-specific evidence in
detail in Section
III.

These rates are not comparable, and we do not pretend they are. They
come from sources that define failure in different ways: abandonment,
never reaching production, weak ROI, thin adoption, drawn from different
samples and different methods. So we do not average them into one
number. We read them instead as convergent signals, independent reports
pointing the same direction, which is reason enough to look harder. The
aim here is the structural cause, not the exact percentage. Table I lays
out the figures with their definitions and their limits.

\begin{table*}[!t]
\centering
\footnotesize
\renewcommand{\arraystretch}{1.25}
\caption{FAILURE STATISTICS: SOURCES, DEFINITIONS, AND LIMITATIONS}
\begin{tabular}{@{}@{}
  >{\raggedright\arraybackslash}p{(\columnwidth - 8\tabcolsep) * \real{0.1983}}
  >{\raggedright\arraybackslash}p{(\columnwidth - 8\tabcolsep) * \real{0.1761}}
  >{\raggedright\arraybackslash}p{(\columnwidth - 8\tabcolsep) * \real{0.0791}}
  >{\raggedright\arraybackslash}p{(\columnwidth - 8\tabcolsep) * \real{0.2685}}
  >{\raggedright\arraybackslash}p{(\columnwidth - 8\tabcolsep) * \real{0.2780}}@{}}
\toprule
\textbf{Statistic} & \textbf{Source} & \textbf{Year} &
\textbf{Definition of \textquotesingle Failure\textquotesingle{}} &
\textbf{Limitations} \\
\midrule
60\%-85\% big data projects fail & Gartner (as reported in {[}2{]}) &
2015-2017 & Project abandonment or failure to deliver expected business
value & Analyst estimate; methodology undisclosed; revised upward
suggesting initial undercount \\
Data lakes become data swamps without governance & Gartner {[}3{]} &
2014 & \textquotesingle Swamp\textquotesingle{} = data without proper
metadata, governance or consumption strategy & Directional warning;
widely cited in industry; \textquotesingle swamp\textquotesingle{}
metaphor subjective \\
87\% of data science projects fail to reach production & VentureBeat
{[}5{]} & 2019 & Failure to deploy ML models to production environments
& Data science scope (not lake-specific); survey methodology;
self-reported \\
98.8\% invest, 37.8\% data-driven & NewVantage {[}4{]} & 2020 &
Organizational transformation gap; investment vs. cultural adoption &
Fortune 1000 only; self-reported;
\textquotesingle data-driven\textquotesingle{} subjectively defined \\
92\% cite culture as obstacle & NewVantage {[}6{]} & 2022 & Executive
perception of primary barrier to data-driven transformation & Executive
self-report; perception vs. objective measurement; Fortune 1000 bias \\
\bottomrule
\end{tabular}
\end{table*}

Culture is where much of this breaks. NewVantage Partners finds 92\% of
executives naming corporate culture, not technology, as the main
obstacle to becoming data-driven {[}6{]}. Bob Muglia, former CEO of
Snowflake, put it more bluntly: he could not find a happy Hadoop
customer, and reckoned fewer than twenty had truly tamed it {[}7{]}. The
money went in. For most, the value did not come out.

What ties these outcomes together, we argue, is a single mechanism.
Governance decisions get deferred, and the cost compounds. We call it
Governance Debt.

Still, this gap deserves a more direct conversation about causes and
about what, if anything, we have learned over the past fifteen years.
The Data Lake promise was bold, but the results have often been
sobering, and the community has not always addressed that contrast
openly. Therefore, a few basic questions arise. Some concern the concept
itself and others concern how it has been executed in real
organizations. After years of costly trials, it is worth asking whether
we have learned meaningful lessons, or whether the same issues are
resurfacing under newer labels such as Lakehouse and Data Mesh.

This work presents a critical retrospective of the Data Lake paradigm.
We focus on recurring failure patterns and the drivers behind them. We
adopt this perspective because the literature often misses the same gap
we observe in practice. Technical surveys tend to enumerate
architectures without challenging core premises {[}8{]}, {[}9{]}.
Vendor-oriented accounts, in contrast, often foreground tooling and
engineering remedies, with limited attention to governance and
operating-model prerequisites.

This retrospective is guided by a set of uncomfortable questions. First,
why do Data Lake projects still fail at high rates despite more than a
decade of experience? Many of the failure patterns reported in 2015 are
still visible in 2025. In practice, organizations that struggled with
first-generation lakes often launch a second attempt with updated
technologies but largely unchanged ways of working, and the results
remain similar. Second, was ``schema-on-read'' ever a realistic model
for enterprise data management at scale? Gartner warned that ``through
2018, 80\% of data lakes will not include effective metadata management
capabilities, making them inefficient'' {[}10{]}. Later industry
evidence remains consistent with this concern, and the underlying issue
persists. Third, why are these failures still framed mainly as
technology problems rather than governance and organizational problems?
Each technology wave, from Hadoop to Spark, then cloud-native stacks and
modern table formats, has been promoted as a reset. Yet recurring
failure patterns continue to surface. Finally, do newer paradigms such
as Data Lakehouse and Data Mesh represent genuine progress, or do they
largely repackage similar ideas? When we examine the discourse around
these paradigms, we find strong echoes of the promises made about Data
Lakes fifteen years ago.

This paper makes five contributions that clarify how Data Lakes evolved
and what their track record implies for modern data architectures:

\begin{enumerate}
\def\labelenumi{\arabic{enumi}.}
\item
  a critical synthesis of fifteen years of Data Lake promises and
  outcomes, including whether the successor paradigms, Data Lakehouse and
  Data Mesh, absorbed the lessons, grounded in 64 documented sources and
  in a primary practitioner catalogue of close to five hundred field
  reality checks;
\item
  the Seven Deadly Sins, a taxonomy of the recurring organizational
  anti-patterns behind Data Lake failure;
\item
  Governance Debt as the central explanatory mechanism, with Governance
  Gravity as its corollary, extended by two field-driven components the
  literature under-reports, operational debt and engineering-discipline
  debt;
\item
  a reading of these mechanisms from a Moroccan and West African
  emerging-market vantage that is largely absent from the existing record
  and that amplifies several of the anti-patterns;
\item
  practical diagnostic and intervention tools for practitioners,
  presented as unvalidated: a Reality Check Framework, a Stage-Based
  Intervention Matrix, and a research agenda for empirical validation.
\end{enumerate}

These five are not equal in weight. The analytical core is the taxonomy,
Governance Debt, and Governance Gravity, which we defend against both the
documented record and the field catalogue. The practitioner catalogue
and the emerging-market reading are primary evidence and context. The
diagnostic tools, the Governance Debt Assessment rubric, the Reality
Check Framework, and the Stage-Based Intervention Matrix, are working
instruments, offered to practitioners and labelled plainly as
unvalidated. Weigh the conceptual core on its own, apart from the toolbox
we build on it.

The remainder of this paper is organized as follows. Section II reviews
related work and positions the contribution. Section III describes our
methodology, including the practitioner reality-check catalogue. Section
IV deconstructs the original promises. Section V presents the
anti-patterns framework. Section VI introduces governance debt and the
causal chain to reversion. Section VII examines successor paradigms.
Section VIII reports the field evidence drawn from the catalogue. Section
IX situates the analysis in the Moroccan and African context. Section X
presents the Reality Check Framework, the Stage-Based Intervention
Matrix, and the research agenda. Section XI concludes.

\section{RELATED WORK AND POSITIONING}

Two bodies of work bear on this paper: the technical literature on data
lakes themselves, and the wider literature on data governance, technical
debt, and architectural evolution that we draw on to explain failure. We
review each, then state where our contribution sits.

\emph{Data lake systems and architecture.} Data lakes have been surveyed
extensively from a systems standpoint. Hai, Geisler, and Quix map the
functions and systems of data lakes and the design space they occupy
{[}8{]}, while Sawadogo and Darmont synthesize data lake architectures
and place metadata management at the center of the problem {[}9{]}. This
line of work is rigorous on storage zones, ingestion, metadata models,
and query layers. What it brackets, by design, is why lakes fail in
organizations that already hold the technology, which is the question we
pursue.

\emph{The data swamp and the governance-failure literature.} The ``data
swamp'' has been a recognized degradation mode since Gartner's early
warning {[}3{]} and Dixon's own revisiting of the concept {[}13{]}. A
large practitioner literature describes it, cataloguing symptoms such as
undocumented data, broken lineage, and eroded trust, and prescribing
catalogs and lineage tooling in response {[}20{]}, {[}22{]}, {[}25{]}.
These accounts are useful but mostly descriptive and single-generation.
They rarely supply an explanatory mechanism that survives across
technology waves, and they seldom connect the failure to the operating
model that produced it.

\emph{Technical debt and the debt metaphor.} Our central construct
extends the technical-debt tradition. Cunningham introduced technical
debt to name the compounding cost of expedient short-term decisions
{[}34{]}, a lineage that runs through the ``code smells'' and
anti-pattern literature {[}18{]}, {[}19{]} and through industry estimates
of debt's drag on delivery {[}35{]}. We argue that deferred governance
behaves like deferred engineering, hence Governance Debt, and we separate
our notion of governance gravity from McCrory's data gravity {[}46{]}:
data gravity is physical, the mass of data attracting compute and
services, whereas governance gravity is organizational, the pull of
institutional trust and semantic contracts back toward the warehouse.
Our own prior work informs two of the debt components, metadata quality
as the determinant of usability \cite{r65} and duplication as a recurring
but tractable failure mode \cite{r66}, as does the case for designing
privacy controls in from the start rather than retrofitting them
\cite{r67}.

\emph{Successor paradigms and migration.} The architectures proposed to
supersede the lake carry their own foundational statements, notably
Dehghani on Data Mesh {[}56{]}, accompanied by a migration literature on
managing legacy transitions: the Strangler Fig pattern {[}47{]}, data
contracts {[}48{]}, and the reliability principles of {[}49{]}. This work
describes target designs and transition tactics well. What it does not
offer is an evaluative, cross-generational account of whether the
organizational failures were actually resolved, which we take up directly
in Section VII.

\emph{The gap, and our position.} Across these strands, one account is
missing: a sustained, cross-generational explanation of why data lakes
fail in organizational rather than technical terms, grounded in
first-hand delivery and written from outside the North American and
European enterprise. That culture outweighs technology is itself well
documented {[}51{]}, but it is rarely tied to a concrete debt mechanism
or to a primary practitioner record. That gap is where this paper sits.
Much of our apparatus, the Seven Deadly Sins and Governance Debt,
relabels and synthesizes patterns that practitioners and analysts had
already described in pieces, and we say so plainly. Two things here are
new. One is the practitioner catalogue, a primary qualitative corpus that
tests the synthesized framework against fifteen years on the ground and
brings out two failure dimensions the surveys skip, operational and
engineering-discipline debt. The other is the emerging-market reading,
which shows how the same universal anti-patterns are amplified by the
talent, regulatory, and vendor conditions of Moroccan and African
enterprises. The contribution is the grounding and the vantage point, not
the coining.

\section{METHODOLOGY}

In this study, we adopt an industry-informed critical retrospective
methodology that draws on two complementary bodies of evidence. The
first is the documented record: academic literature, analyst reports,
and practitioner accounts. The second is a primary practitioner corpus:
a catalogue of field reality checks that the author recorded across
fifteen years of enterprise Data Lake delivery. The documented record
lets us reconstruct the industry narrative and test its claims. The
practitioner corpus checks that narrative against what actually happened
on the ground. We pair them because the literature has a known blind
spot. Failed Data Lakes rarely show up in peer-reviewed work: few
organizations publish their own post-mortems, and the incentives run the
other way. The people who do see the failures, up close and repeatedly,
are consultants, analysts, and practitioners, in engagements, surveys,
and post-mortems. Lean on academic sources alone and the picture comes
out incomplete, maybe even misleading.

\subsection{Source Taxonomy}

We systematically collected and analyzed sources across three
complementary tiers.

\textbf{Tier 1} covers academic peer-reviewed literature. The search
spanned IEEE Xplore, the ACM Digital Library, ScienceDirect, and
Springer, with journals including IEEE Transactions on
Knowledge and Data Engineering, Information Systems, and Data \&
Knowledge Engineering. Two surveys in particular guided our work, Hai et
al. {[}8{]} and Sawadogo and Darmont {[}9{]}.

\textbf{Tier 2} draws on industry reports and white papers. We reviewed
material from McKinsey, BCG, Deloitte, Accenture, and PwC. We also
included reports from Gartner, Forrester, and IDC. Among these, the
NewVantage Partners annual Big Data and AI Executive Survey presented in
{[}4{]} and {[}6{]} deserves mention, since it has tracked Fortune 1000
companies since 2012 and documents recurring adoption challenges. The
McKinsey reports in {[}11{]} and {[}12{]}
offer a complementary view, focusing on how organizations capture value
at the market level.

\textbf{Tier 3} covers practitioner accounts and foundational texts. Here
we trace the concept to its origins through James
Dixon\textquotesingle s seminal 2010 blog post {[}1{]} and his 2014
clarification on data lakes {[}13{]}, and we draw on
practitioner perspectives from technical conferences and post-mortem
analyses.

In total, we analyzed 64 sources: 16 peer-reviewed
academic publications and books, 30 industry analyst reports and
surveys, and 18 practitioner accounts and vendor documents. Sources were
identified through systematic search across academic databases
(IEEE Xplore, ACM Digital Library, Google Scholar) and industry
repositories (Gartner, Forrester, McKinsey) using keywords including
\textquotesingle data lake failure,\textquotesingle{}
\textquotesingle data swamp,\textquotesingle{} \textquotesingle big data
governance,\textquotesingle{} and \textquotesingle data lake
implementation.\textquotesingle{} To be included, a source had to
address Data Lake implementation outcomes, governance challenges, or
failure patterns, and to offer documented evidence or a structured
methodology. We excluded three types of source: (i)
those focusing on technical architecture without organizational context,
(ii) those predating the Data Lake concept (before 2010), and (iii)
those lacking identifiable authorship or institutional affiliation.

After the collection process, we moved to coding failure patterns using
an inductive approach. A first round of open coding identified 23
distinct failure modes. We then consolidated these into seven thematic
categories through axial coding, grouping them by shared root causes. To
reduce bias, we triangulated quantitative claims across at least two
independent sources whenever possible. Moreover, we flagged claims that
rely on just one source. Vendor documentation required specific
attention, since it is often promotional, we cross-referenced it against
independent analyst reports. In Table II, we summarize the full search
protocol.

\begin{table*}[!t]
\centering
\footnotesize
\renewcommand{\arraystretch}{1.25}
\caption{SEARCH PROTOCOL SUMMARY}
\begin{tabular}{@{}@{}
  >{\raggedright\arraybackslash}p{(\columnwidth - 2\tabcolsep) * \real{0.1817}}
  >{\raggedright\arraybackslash}p{(\columnwidth - 2\tabcolsep) * \real{0.8183}}@{}}
\toprule
\textbf{Element} & \textbf{Specification} \\
\midrule
Databases & IEEE Xplore, ACM Digital Library, Google Scholar, Gartner,
Forrester, McKinsey Digital \\
Search period & 2010-2025 (publication date); searches conducted
September-December 2025 \\
Keywords & \textquotesingle data lake failure\textquotesingle,
\textquotesingle data swamp\textquotesingle, \textquotesingle big data
governance\textquotesingle, \textquotesingle data lake
implementation\textquotesingle, \textquotesingle data lake
challenges\textquotesingle, \textquotesingle data governance
maturity\textquotesingle, \textquotesingle data warehouse
migration\textquotesingle{} \\
Inclusion criteria & Addresses implementation outcomes, governance
challenges, or failure patterns; documented evidence or structured
methodology; identifiable authorship/affiliation \\
Exclusion criteria & Technical architecture only (no organizational
context); pre-2010; anonymous/unaffiliated authors; product marketing
without evidence \\
Coding method & Inductive: open coding (23 failure modes) $\rightarrow$ axial coding
(7 thematic categories based on shared root causes) \\
Bias mitigation & Triangulation across $\geq$2 sources for quantitative
claims; vendor content cross-referenced with independent analysts;
single-source claims explicitly attributed \\
\bottomrule
\end{tabular}
\end{table*}

\subsection{The Practitioner Catalogue}

The second body of evidence is a primary practitioner corpus. Over
fifteen years of designing, delivering, and rescuing enterprise Data
Lake platforms in banking, consumer finance, and telecommunications
across Morocco and West Africa, the author kept a running catalogue of
dated field ``reality checks,'' short statements of recurring failure
modes observed on the ground. It grew to close to five hundred entries.
We treat it as a qualitative corpus, not a source of metrics: deriving
failure rates from confidential engagements would present recollection as
measurement. We consolidated the entries by removing duplicates and
near-duplicates, then coded them thematically against the framework, into
the sixteen clusters provided as supplementary material.

The coding is itself a result. Built from practice, with no eye on the
analyst literature, the catalogue lands almost entirely on the same
anti-patterns that literature describes, and it adds two dimensions the
literature under-reports, operational and engineering-discipline debt.
That convergence is the validation I rely on instead of contestable
statistics. I was also an actor in many of these situations, which grants
access an outside reviewer lacks but brings selection and hindsight
effects; I address these in Section X and triangulate against the
documented record throughout.

\subsection{Temporal Scope and Periodization}

\begin{figure*}[!t]
\centering
\includegraphics[width=\textwidth]{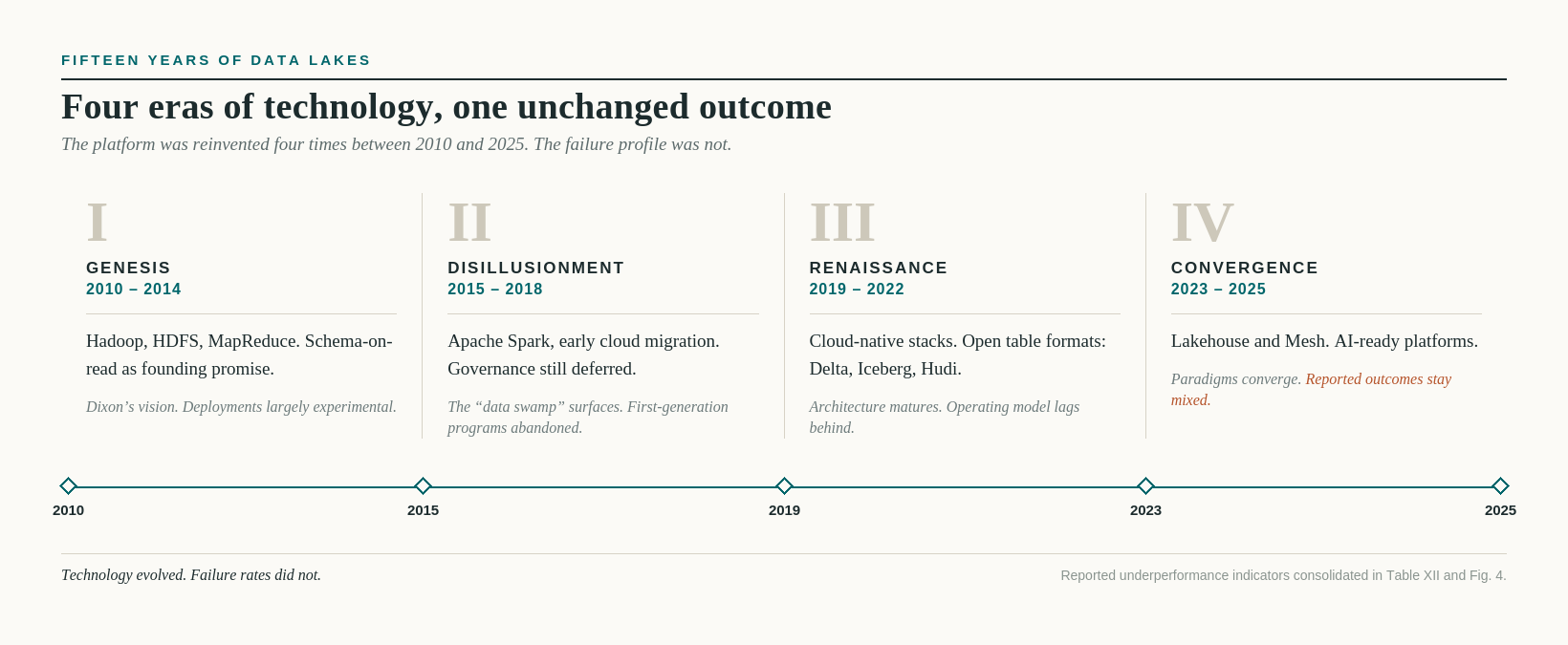}
\caption{Four eras of Data Lake technology (2010 to 2025). The platform was reinvented four times while the reported failure profile did not move. Underperformance indicators are consolidated in Table~\ref{tab:flat} and Fig.~\ref{fig:flat}.}
\label{fig:eras}
\end{figure*}

Our analysis spans 2010 to 2025, divided into four periods defined by
the dominant technology, the prevailing discourse around Data Lakes, and
the observed outcomes. Fig.~\ref{fig:eras} shows the timeline.

\begin{enumerate}
\def\labelenumi{\arabic{enumi}.}
\item
  In the Genesis phase, from 2010 to 2014, Dixon introduced the concept
  and most implementations were Hadoop-centric. Expectations ran high,
  yet most deployments stayed experimental.
\item
  In the Disillusionment phase, from 2015 to 2018, the data swamp
  problem came to the fore. Gartner raised its failure-rate estimate to
  85\% {[}2{]}, and many organizations abandoned their first-generation
  projects.
\item
  In the Renaissance phase, from 2019 to 2022, cloud-native solutions
  matured, open table formats emerged (Delta Lake, Apache Iceberg, and
  Apache Hudi), and architectural practice improved.
\item
  In the Convergence phase, from 2023 to 2025, the Data Lakehouse gained
  traction, Data Mesh became a real option, and AI/ML requirements began
  to shape platform design. Outcomes nonetheless remained mixed: BCG
  reports that only 35\% of digital transformation initiatives meet their
  objectives {[}14{]}.
\end{enumerate}

\subsection{Analytical Framework}

We examine each claim about Data Lakes through three questions: what the
foundational texts actually promised, what assumptions those promises
left implicit, and what the evidence then shows. We label a promise as
\textquotesingle failed\textquotesingle{} only when at least two of our
three source tiers agree.

\subsection{Limitations}

This is a critical retrospective synthesis, not a controlled empirical
study: it identifies recurring patterns across heterogeneous sources and
proposes explanatory frameworks for later testing rather than
establishing causal relationships through experiment. The evidence
therefore carries the known biases of its sources, survivorship in
failure reporting, promotional slant in vendor material, and greater
uncertainty for the most recent period, which we mitigate by
triangulating across tiers and flagging single-source claims. The
instruments we propose, the GDAM rubric, the swamp indicators, and the
governance gravity reasoning, are testable hypotheses, not validated
measurement tools. We develop these points, and the threats to validity,
in Section X.

\begin{table*}[!t]
\centering
\footnotesize
\renewcommand{\arraystretch}{1.25}
\caption{SOURCE TAXONOMY AND EVIDENCE BASE}
\begin{tabular}{@{}@{}
  >{\raggedright\arraybackslash}p{(\columnwidth - 6\tabcolsep) * \real{0.2126}}
  >{\raggedright\arraybackslash}p{(\columnwidth - 6\tabcolsep) * \real{0.2399}}
  >{\raggedright\arraybackslash}p{(\columnwidth - 6\tabcolsep) * \real{0.2618}}
  >{\raggedright\arraybackslash}p{(\columnwidth - 6\tabcolsep) * \real{0.2856}}@{}}
\toprule
\textbf{Tier} & \textbf{Source Type} & \textbf{Examples} &
\textbf{Evidence Type} \\
\midrule
Tier 1: Academic & Peer-reviewed journals, conference proceedings &
IEEE, ACM, VLDB, Information Systems & Architecture patterns, technical
frameworks \\
Tier 2: Industry & Analyst reports, surveys, market research & Gartner,
McKinsey, Forrester, NewVantage & Adoption rates, failure statistics,
market trends \\
Tier 3: Practitioner & Case studies, post-mortems, tech blogs &
Engineering blogs, conference talks, retrospectives & Implementation
challenges, lessons learned \\
\bottomrule
\end{tabular}
\end{table*}

\section{THE ORIGINAL PROMISES: DECONSTRUCTING THE DATA LAKE VISION}

In October 2010, James Dixon, then CTO of Pentaho, published the blog
post that introduced the Data Lake concept {[}1{]}, an idea that would
shape enterprise data architecture for the next fifteen years. His
metaphor was simple: where a warehouse holds ``bottled water,''
structured and packaged, the lake holds water in a more natural state.
The image caught on quickly, and vendors and analysts adopted it in
turn. PwC went as far as claiming that Data Lakes could ``put an end to
data silos'' {[}15{]}, and McKinsey made comparable claims, assuming
that big data analytics would bring major returns {[}11{]}. The
assumptions behind these promises are worth questioning. In the sections
that follow we examine each promise in turn, what it assumed and whether
reality bore it out, with Table IV giving an overview of the analysis we
then detail.

\subsection{Promise 1: The End of Data Silos}

The first promise was the end of silos. Data Lakes, the argument went,
would gather everything in one place and dissolve the fragmentation that
plagued enterprise data, and PwC went so far as to say they could put an
end to data silos {[}15{]}. The claim only holds if silos are a storage
problem: if teams fail to share data because they lack a common place to
put it, and if, given that place, they will share willingly. Neither
survives contact with how organizations actually behave. Silos are
political and cultural before they are technical, and a shared bucket
does nothing to change the incentives that created them. The evidence
bears this out. NewVantage Partners finds 92\% of executives naming
corporate culture, not technology, as the main barrier {[}6{]}, and
McKinsey reports that in manufacturing, the public sector, and
healthcare, organizations have captured less than 30\% of the value
analytics could deliver {[}12{]}.

\subsection{Promise 2: Schema-on-Read Flexibility}

Schema-on-read was the second promise. Where a warehouse forces you to
define structure before loading, the lake would let you ingest raw and
impose schema only at query time. The appeal was real, but it rested on
two shaky beliefs: that deferring schema work reduces the total work,
and that a future user will still understand raw data years later
without the context that produced it. Deferral does not remove the
burden, it relocates it onto everyone downstream, and Gartner saw the
failure mode coming when it predicted that through 2018, 80\% of data
lakes would not include effective metadata management capabilities
{[}10{]}. The field bore it out: schema-on-read quietly became
schema-never, and data that no one could interpret became data no one
could use.

\subsection{Promise 3: Store Everything, Analyze Later}

The third promise
was to store everything and analyze later, on the theory that today's
exhaust might be tomorrow's insight. This assumed that raw data stays
usable without active curation and that storage was the dominant cost,
which quietly set aside the costs of governance, security, and
compliance and, underneath, mistook more data for more value. Gartner's
2014 warning that ungoverned lakes would become data swamps {[}3{]}
proved exactly right: lakes filled with data that no one could find,
understand, or trust became the defining failure of the era.

\subsection{Promise 4: Democratization of Data Access}

The fourth promise was democratization: analysts and scientists reaching
raw data directly, without waiting on IT, and self-service analytics
flourishing as a result. The premise was that access was the binding
constraint, which overlooks skills, context, and literacy, and forgets
that IT gatekeeping often exists for good reasons, namely quality,
security, and compliance. The numbers tell the rest. DataCamp's 2024
report finds only 28\% of organizations have reached data literacy even
as 83\% of leaders call it critical {[}16{]}. Democratization in
practice came to mean that everyone could reach the lake while only a
few knew how to swim in it.

\subsection{Promise 5: Cost Reduction Through Commodity Infrastructure}

The fifth promise was cost: Hadoop on commodity hardware would undercut
proprietary warehouse appliances, and the saving looked self-evident. It
looked that way only because the comparison stopped at infrastructure.
It ignored total cost of ownership, assumed the organization had the
skills to run distributed systems, and treated open source as free while
enterprise needs quietly added their own bill. Gartner again called it,
predicting that through 2018, 70\% of Hadoop deployments would fail to
meet cost savings and revenue generation objectives owing to skills and
integration challenges {[}10{]}, and experience confirmed that
integration, training, and change management routinely cost more than
the software, sometimes by a wide margin {[}14{]}.

\subsection{Promise 6: Single Source of Truth}

The last promise was a single source of truth: one place, one version,
no more conflicting copies. This conflated two different things, putting
data in one place and making data consistent, and assumed that physical
centralization would resolve a logical problem. It does not. Different
systems define customer or revenue differently, and storing those
definitions side by side leaves the disagreement intact. NewVantage
Partners finds only 37\% of organizations reporting improved data
quality {[}17{]}; more often the lake became one more silo rather than
the end of them.

\subsection{The Promise Pyramid: A Synthesis}

When we look at these promises together, a pattern emerges:
technological solutionism. In other words, the belief that technology
alone can solve organizational and human problems. The promises also
formed a kind of pyramid, where each layer depended on the one below.
Cost reduction made it possible to store everything. Storing everything
allowed for schema flexibility. Schema flexibility enabled
democratization. And democratization would end silos and create a single
source of truth. But if any layer at the base failed, the whole
structure would collapse.

So what did the promises miss? That managing data is a sociotechnical
problem, not a technical one. Technology hands you capabilities. Value
takes more: organizational alignment, skills, working governance, a
culture ready to use any of it. The Data Lake sold a technical fix to
problems that were mostly about people and process.

\begin{table*}[!t]
\centering
\footnotesize
\renewcommand{\arraystretch}{1.25}
\caption{ORIGINAL PROMISES VS. IMPLICIT ASSUMPTIONS VS. REALITY}
\begin{tabular}{@{}@{}
  >{\raggedright\arraybackslash}p{(\columnwidth - 4\tabcolsep) * \real{0.3179}}
  >{\raggedright\arraybackslash}p{(\columnwidth - 4\tabcolsep) * \real{0.3380}}
  >{\raggedright\arraybackslash}p{(\columnwidth - 4\tabcolsep) * \real{0.3442}}@{}}
\toprule
\textbf{Original Promise} & \textbf{Hidden Assumption} & \textbf{What
Actually Happened} \\
\midrule
Schema Flexibility & Users can impose schema discipline at consumption
time & Schema-on-read became schema-never; 80\% lack metadata management
{[}10{]} \\
Data Democratization & Access would be matched by data literacy and
analytical skills & Only 28\% data literacy vs 83\% say critical
{[}16{]}; access $\neq$ usability \\
Cost Reduction & Hardware savings would not be consumed by skills and
integration costs & Integration costs often exceed software costs; TCO
frequently exceeds initial estimates \\
Scalability & Organizations would scale data management practices
alongside volume & Governance debt compounds; larger scale = larger
ungoverned swamp \\
Agility & Faster ingest would translate to faster analytics delivery &
Substantial time spent on data quality remediation; discovery
bottleneck \\
Unified Repository & Centralization would naturally lead to consistency
and integration & Departments manage independently; duplicates
proliferate; no ownership \\
\bottomrule
\end{tabular}
\end{table*}

\section{WHAT WENT WRONG: THE SEVEN DEADLY SINS OF DATA LAKES}

\begin{figure*}[!t]
\centering
\includegraphics[width=0.92\textwidth]{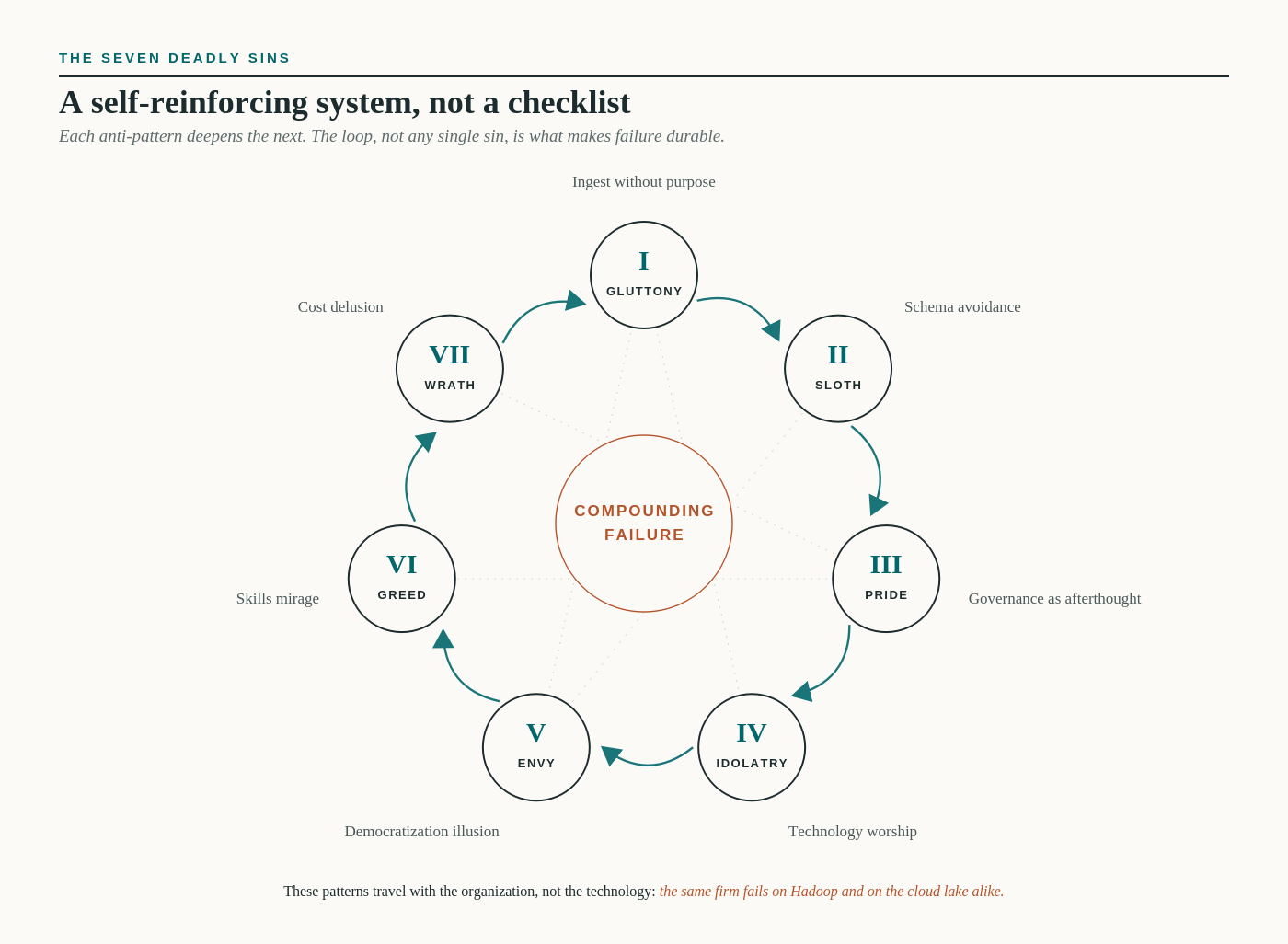}
\caption{The Seven Deadly Sins as a self-reinforcing system. Each anti-pattern deepens the next; the loop, not any single sin, is what makes failure durable across technology generations.}
\label{fig:sins}
\end{figure*}

The promises failed in patterns, not at random. Reading across the
industry reports, academic studies, and practitioner accounts, the same
seven anti-patterns surface again and again. We call them the Seven
Deadly Sins, and they account for most of the documented Data Lake
failures (Table V). They are not independent. Each one feeds the next,
and it is the compounding loop, more than any single mistake, that
shrugs off purely technological fixes (Fig.~\ref{fig:sins}).

\textbf{Terminological Note, Taxonomy vs Mnemonic:} The "Seven Deadly
Sins" label requires methodological clarification. In software
engineering and information systems research, taxonomies of recurring
problems commonly employ evocative mnemonic labels to facilitate
practitioner recognition and organizational discourse, examples include
"code smells" {[}18{]}, "anti-patterns" {[}19{]}, and "architecture
smells" in technical debt literature. We adopt this convention: the
taxonomy presented here constitutes a structured classification of
\emph{socio-technical failure modes}, not moral judgments. The "sins"
are organizational tendencies arising from structural incentives,
information asymmetries, and bounded rationality, not individual
failings. Each pattern represents a probabilistic tendency that
organizations exhibit to varying degrees depending on context; the
taxonomy\textquotesingle s value lies in providing a shared vocabulary
for diagnosing and discussing governance challenges. We retain the
mnemonic label for its practical utility in organizational communication
while emphasizing that the underlying analysis is grounded in documented
patterns across multiple independent sources.

Before examining these patterns, we establish an operational definition.
A \emph{Data Swamp} is operationally defined as a Data Lake exhibiting
three or more of the following measurable indicators: (1) absence of
assigned dataset ownership for \textgreater50\% of assets; (2) metadata
completeness below 40\% (fewer than two-fifths of required fields
documented); (3) more than 40\% of datasets unused in the past 90 days;
(4) undocumented or untraceable data lineage for critical pipelines; (5)
duplicate datasets exceeding 20\% of total assets; or (6) no automated
quality validation on ingestion pipelines. This definition transforms
the metaphorical "swamp" into an assessable state, enabling
organizations to diagnose degradation before it becomes irreversible.
These thresholds represent pragmatic heuristics derived from observed
patterns in practitioner accounts and industry surveys (e.g., 90 days as
a common governance monitoring window, 40\% as a tipping point below
which discoverability collapses) rather than empirically validated
cutoffs. Calibration should consider organizational context:
data-intensive industries (finance, healthcare, pharmaceuticals) should
apply stricter thresholds (e.g., 30\% rather than 40\% for metadata
completeness), while exploratory or research environments may tolerate
higher variance. Sensitivity analysis suggests that the diagnostic value
lies not in absolute thresholds but in tracking trajectory:
deteriorating metrics over consecutive quarters signal active swamp
formation regardless of current absolute values. A practical approach
treats threshold breaches as leading indicators warranting investigation
rather than definitive diagnoses, the pattern of multiple simultaneous
breaches is more diagnostic than any single threshold.

Operationalizing these indicators requires
instrumentation that many organizations lack. Usage tracking draws on
query logs, SQL audit trails, and BI platform consumption metrics.
Ownership and lineage assessment leverages data catalog metadata,
lineage tool dependency graphs, and stewardship assignments in
governance platforms. Duplication detection employs automated profiling
tools that compute schema similarity and content hashing across
datasets. Metadata completeness is measured against catalog field
requirements. Validation coverage derives from CI/CD pipeline
configurations and data quality tool deployments. Organizations lacking
instrumentation for these measurements face a meta-problem: they cannot
diagnose swamp status because the diagnostic infrastructure itself was
never implemented, a common manifestation of the
\textquotesingle governance as afterthought\textquotesingle{} pattern.
The following matrix provides concrete measurement guidance for
enterprise environments:

Each swamp indicator can be made concrete with a
formula, a data source, and a heuristic threshold that each organization
should calibrate rather than adopt verbatim: \emph{(1) Ownership Coverage Rate},
Formula: (datasets with assigned steward) / (total datasets) $\times$ 100.
Source: data catalog stewardship field (Collibra, Alation, Atlan, Apache
Atlas). Threshold: \textless50\% indicates swamp risk (heuristic).
\emph{(2) Metadata Completeness Index}, Formula: $\Sigma$(populated required
fields) / $\Sigma$(total required fields) $\times$ 100, aggregated across all
datasets. Source: catalog field completion statistics. Threshold:
\textless40\% indicates swamp risk (heuristic; data-intensive industries
should use \textless30\%). \emph{(3) Dataset Dormancy Rate}, Formula:
(datasets with zero queries in 90 days) / (total datasets) $\times$ 100.
Source: query logs (Snowflake ACCOUNT\_USAGE, Databricks query history,
BigQuery INFORMATION\_SCHEMA.JOBS). Threshold: \textgreater40\%
indicates swamp risk (heuristic; 90 days is standard governance window).
\emph{(4) Lineage Coverage Gap}, Formula: (critical pipelines with
documented lineage) / (total critical pipelines) $\times$ 100. Source: lineage
tool (OpenLineage, Marquez, Unity Catalog). Threshold: \textless100\%
for regulatory/financial pipelines indicates swamp risk; \textless80\%
for operational pipelines (heuristic). \emph{(5) Duplication Ratio},
Formula: (duplicate dataset pairs identified by schema similarity
\textgreater90\% OR content hash match) / (total datasets) $\times$ 100.
Source: profiling tools (Great Expectations, Soda, Monte Carlo).
Threshold: \textgreater20\% indicates swamp risk (heuristic). \emph{(6)
Validation Coverage Rate}, Formula: (production pipelines with automated
quality gates) / (total production pipelines) $\times$ 100. Source: CI/CD
configurations, dbt test coverage reports. Threshold: \textless100\% for
production pipelines indicates swamp risk (heuristic; any ungated
pipeline is a quality liability).

A critical distinction separates an \emph{immature lake} from a true
\emph{swamp}, and it matters because it answers the obvious objection
that the analysis might brand any struggling initiative a swamp. An
immature lake, typically Stage 1 or 2 of the degradation pathway with a
mostly Partial to Established profile, still has governance gaps, but it
retains the institutional knowledge to understand its data, remediation
costs proportional to its capacity, the organizational will to invest,
and assets whose residual value exceeds the cost of cleanup. A swamp,
Stage 3 to 5 with a mostly Absent to Ad hoc profile, is the tipping point
where those conditions invert: the people who understood the data have
left, remediation costs now exceed the data's residual value, governance
fatigue has set in, and debt compounds faster than any realistic
remediation can keep up. The practical test is the ratio of remediation
effort to capacity. An organization that could close its gaps within a
year or so using existing resources has an immature lake; one that would
need a multi-year effort beyond its budget or its people has a swamp.
This is why some troubled initiatives recover and others spiral into
abandonment: what separates them is not the current severity of symptoms
but whether the cost of fixing them still fits within organizational
capacity.

\subsection{Sin 1: Ingest Without Purpose (Gluttony)}

The first sin is gluttony: ingesting data indiscriminately and treating
storage capacity as a virtue in itself. ``Store everything, analyze
later'' sounds prudent, but it mostly defers the hard question of which
data actually serves a goal, so application logs, sensor feeds, legacy
exports, and social streams pour in with no articulated purpose. The
lake then fills with data that nobody understands, maintains, or uses,
which is exactly how the ``data swamp'' is born, a term that surfaced
around 2015 for repositories of undocumented, unvalidated content
{[}20{]}. The root of the problem is organizational rather than
technical: when each function curates data its own way, sales,
marketing, and finance end up with three incompatible versions of the
same customer {[}21{]}, records are duplicated with conflicting values,
and accountability evaporates the moment something breaks. None of this
was unforeseen. Gartner had warned in 2014 that ungoverned lakes would
degrade into swamps {[}3{]}, and the surveys since have only repeated the
diagnosis: deprived of governance, lake data drifts into inconsistency,
inaccuracy, and incompleteness, and the insights drawn from it stop being
trustworthy {[}22{]}.

\subsection{Sin 2: Schema Avoidance (Sloth)}

If gluttony fills the lake, sloth leaves it illegible. ``Schema-on-read''
was meant to allow different schemas for different uses; in practice it
was read as permission to skip schema work altogether, and the deferred
became the never. Data went in raw and stayed raw. The cost lands on
everyone downstream, since each consumer must reconstruct meaning,
structure, and quality on their own, when they can interpret the data at
all. The knowledge needed to do so usually lives only in the heads of the
original source-system owners, and it erodes as those people move on.
Metadata is not kept current, so it becomes hard to judge the context,
quality, or even the relevance of a dataset {[}23{]}, and users cannot
tell what a file holds, where it came from, or whether it still matters
{[}24{]}. Gartner's forecast that eighty percent of lakes would lack
effective metadata management {[}10{]} reads, in hindsight, as
description rather than prediction: analysts and scientists end up
spending their time discovering, cleaning, and integrating data instead
of analyzing it {[}23{]}, the exact inverse of the productivity the lake
was sold on.

\subsection{Sin 3: Governance as Afterthought (Pride)}

Pride is the conviction that governance can wait, that order can be
imposed on chaos after the fact, and that a capable enough technical team
can build its way past the need for organizational discipline. It treats
governance as bureaucratic overhead to be minimized rather than as
foundation. The trouble is that every later remediation then inherits an
impossible task: to reconstruct, retrospectively, the meaning of data
that was never documented. Organizations without a governance framework
struggle with accuracy, compliance, and standardization until their lakes
``swampify'' {[}25{]}, and the security exposure grows in parallel, as
absent access controls let the wrong people reach sensitive data and turn
gaps into breaches {[}26{]}. The evidence that this is a matter of culture
rather than tooling is by now overwhelming. NewVantage Partners has found,
year after year, that culture rather than technology is the main obstacle
to becoming data-driven, with ninety-two percent of executives saying so
in 2022 {[}6{]}, and the same panels report that for five straight years
the biggest impediment to data initiatives has been cultural, not
technical {[}27{]}.

\subsection{Sin 4: Technology Worship (Idolatry)}

Idolatry is the faith that the next platform will redeem the last. Each
wave, Hadoop, Spark, cloud-native stacks, then Delta Lake and Iceberg,
arrives carrying the hope that \emph{this} one will finally succeed,
without anyone asking whether the previous failure was ever technological
to begin with. So organizations cycle through migrations that consume
enormous resources and leave the underlying process and ownership
problems untouched. The Blackboard team described the trap with unusual
candor: a heavy DevOps lift, data pulled out of relational databases,
pushed through Hadoop, then sharded back into relational databases, an
architecture that simply felt cumbersome end to end {[}28{]}, and, asked
whether they had been pulled in by the hype, the lead admitted as much.
The wider record agrees. Firms without an engineering culture failed at
markedly higher rates than technology-native startups {[}29{]}, and as
Gartner's Nick Heudecker put it, the binding constraint is integration,
linking sources to produce an outcome, not aggregation, yet organizations
keep adopting lakes in the belief that pooling data will by itself yield
insight {[}7{]}.

\subsection{Sin 5: The Democratization Illusion (Envy)}

Envy is the wish to be Google or Amazon without their organizational DNA,
to replicate the capabilities of data-native firms by buying their
posture rather than building their discipline. It collapses two different
things, access and usability, into one, on the assumption that opening
the raw data to everyone will let business users perform sophisticated
analysis. What follows is access without competence: people can reach
the lake but cannot work it, searches grow long and fruitless, and the
difficulty of finding anything becomes itself a sign the lake is turning
into a swamp {[}25{]}. The literacy gap is the hard limit. Eighty-three
percent of leaders call data literacy essential for every role, yet only
twenty-eight percent of organizations have reached it {[}16{]}, so in
practice everyone can enter the lake but only specialists can swim. The
firms that did succeed did so by attracting and keeping the best people,
who brought not just programming but analytical depth, business judgment,
and problem-solving {[}30{]}, capabilities that no technology
democratizes on its own and that the talent pipeline is still years from
supplying at scale {[}30{]}.

\subsection{Sin 6: The Skills Mirage (Greed)}

Greed, in this setting, is the refusal to pay for the talent the platform
demands. Organizations badly underestimate the specialized skills a lake
needs and assume database administrators and analysts can be retrained in
a hurry, or that a handful of data scientists can serve a whole
enterprise. The result is projects that are understaffed and
under-skilled. Data science roles remain among the hardest to fill, named
as difficult by more than sixty percent of hiring managers in 2024
{[}31{]}, and the shortage reaches well beyond scientists, since scarce
and expensive engineers mean that firms without an engineering culture
fail disproportionately {[}29{]}; IBM put the United States shortfall at a
quarter of a million data professionals, with over forty percent of
companies saying the gap was already hurting their competitiveness
{[}32{]}. Gartner had tied the outcome directly to this cause, predicting
that seventy percent of Hadoop deployments would miss their cost and
revenue goals on skills and integration grounds {[}10{]}. Blackboard
again makes it concrete: trying to roll their own distribution rather than
adopt a packaged one, they ended up dedicating two of ten engineers
purely to wrangling open-source tools, patching bugs, and building
work-arounds {[}28{]}.

\subsection{Sin 7: The Cost Delusion (Wrath)}

The last sin closes the loop. Organizations fixate on infrastructure cost
and stay blind to total cost of ownership, setting ``cheap'' Hadoop
storage against ``expensive'' warehouses while ignoring people,
operations, governance, and the cost of the failures themselves. When the
real bill arrives, the response is often wrath aimed at the technology
team rather than reflection on the planning that produced it. Projects
that looked economically justified turn out far more expensive than
forecast and are cut back or abandoned, because integration, training,
and change management routinely dwarf the software spend {[}14{]}. Those
who chased savings by rolling their own Hadoop discovered that the care
and feeding of the environment far exceeded what they had budgeted
{[}28{]}, and McKinsey warned that leaving the skills gap unaddressed
could cost trillions in lost global output {[}33{]}. The ``free'' label
on open source accelerated adoption and set expectations it could never
meet {[}29{]}: once the specialized talent, the operational complexity,
and the eventual governance remediation are counted, the low-cost promise
turns out to have been the most expensive assumption of all.

\subsection{The Interaction of Sins: Compounding Failures}

These seven sins do not operate in isolation. They form a reinforcing
system where each failure pattern amplifies the others. Ingest Without
Purpose creates volumes of data that Schema Avoidance leaves
undocumented. Governance as Afterthought means no one owns the problem.
Technology Worship leads organizations to seek technical solutions to
these organizational failures. The Democratization Illusion creates
frustrated users who cannot utilize the lake. The Skills Mirage means no
one has the expertise to remediate. The Cost Delusion ensures that when
problems become visible, resources have already been exhausted.

This interaction explains why Data Lake failures are so persistent
despite technological advances. Each generation of technology, Hadoop,
Spark, cloud-native platforms, lakehouses, addresses some technical
limitations of its predecessors but cannot address the organizational
patterns that drove failure in the first place. Organizations that
failed with Hadoop proceed to fail with cloud data lakes in remarkably
similar ways, because the sins travel with the organization, not the
technology.

\begin{table*}[!t]
\centering
\footnotesize
\renewcommand{\arraystretch}{1.25}
\caption{THE SEVEN DEADLY SINS OF DATA LAKES: SUMMARY OF ANTI-PATTERNS}
\begin{tabular}{@{}@{}
  >{\raggedright\arraybackslash}p{(\columnwidth - 6\tabcolsep) * \real{0.1665}}
  >{\raggedright\arraybackslash}p{(\columnwidth - 6\tabcolsep) * \real{0.3078}}
  >{\raggedright\arraybackslash}p{(\columnwidth - 6\tabcolsep) * \real{0.2935}}
  >{\raggedright\arraybackslash}p{(\columnwidth - 6\tabcolsep) * \real{0.2323}}@{}}
\toprule
\textbf{Sin} & \textbf{Anti-Pattern} & \textbf{Key Statistic} &
\textbf{Source} \\
\midrule
Gluttony & Ingest without purpose; store everything without consumption
plan & Risk of degrading into data swamps (widely cited) & Gartner
{[}3{]} \\
Sloth & Schema avoidance; schema-on-read becomes schema-never & 80\%
lack metadata management & Gartner {[}10{]} \\
Pride & Governance as afterthought; belief it can be retrofitted & 92\%
cite culture as main obstacle & NewVantage {[}6{]} \\
Idolatry & Technology worship; each wave embraced as silver bullet &
70\% Hadoop failures: skills/integration & Gartner {[}10{]} \\
Envy & Democratization illusion; access confused with usability & 28\%
data literacy vs 83\% critical & DataCamp {[}16{]} \\
Greed & Skills mirage; underestimating specialized expertise required &
60\%+ cite data science hardest to fill & Industry surveys {[}31{]},
{[}32{]} \\
Wrath & Cost delusion; focus on infrastructure ignoring TCO &
Integration costs exceed software & BCG {[}14{]} \\
\bottomrule
\end{tabular}
\end{table*}

\emph{Note: Statistics derive from surveys conducted 2017-2024 across
enterprise populations (Fortune 1000 for NVP, global enterprises for
Gartner/DataCamp/BCG). Specific years and populations: {[}3{]} Gartner
2017 prediction; {[}10{]} Gartner 2017 governance survey; {[}16{]}
DataCamp 2024 (n=550+ leaders, US/UK); {[}31{]}, {[}32{]} industry
skills surveys 2023-2024; {[}14{]} BCG 2021 digital transformation study
(n=850+ companies).}

\textbf{Evidence Frequency:} The taxonomy emerged from iterative coding
of the 64 sources, with 23 distinct failure modes consolidated into
seven categories by thematic affinity. Most sources touch several
anti-patterns at once, so the categories are not mutually exclusive. Read
qualitatively, the governance-related sins, purposeless ingestion (Sin 1)
and governance as afterthought (Sin 3), are the most frequently cited
across the corpus, consistent with our central thesis, while the
democratization illusion (Sin 5) is the least. Every category cleared our
inclusion floor of at least three independent sources.

\section{THE GOVERNANCE DEBT: A FRAMEWORK FOR UNDERSTANDING ACCUMULATED LIABILITIES}

Look across the seven sins and one thread runs through all of them:
governance keeps getting put off. We give that thread a name, Governance
Debt. It extends the familiar technical-debt metaphor to the decisions
organizations defer about data, and to the way those deferrals compound
over time.

\subsection{From Technical Debt to Governance Debt}

The concept of technical debt, introduced by Ward Cunningham in 1992,
provides a framework for understanding the long-term consequences of
expedient short-term decisions in software development {[}34{]}. Like
financial debt, technical debt accumulates interest over time, the
longer remediation is deferred, the more costly it becomes. McKinsey
reports that CIOs estimate technical debt amounts to 20-40\% of their
technology estate {[}35{]}. Research indicates that 41\% of developer
time is spent addressing bugs, maintenance, and technical debt rather
than new development {[}36{]}, with organizations reporting that
technical debt significantly impacts their ability to deliver new
features {[}37{]}.

We extend this metaphor to data governance, arguing that governance debt
shares the core characteristics of technical debt but exhibits
distinctive properties that make it particularly insidious in Data Lake
contexts. Like technical debt, governance debt represents deferred work
that must eventually be addressed. Unlike technical debt, however,
governance debt involves not just technical remediation but
organizational realignment, cultural change, and the reconstruction of
institutional knowledge that may have been permanently lost.

\subsection{The Components of Governance Debt}

We identify seven distinct components of governance debt that accumulate
in Data Lake environments (Table IX). Five are established in the
governance literature; the last two, operational debt and
engineering-discipline debt, are added here because the field catalogue
of Section VIII shows them to be as lethal as the others yet largely
absent from the analyst record.

\textbf{1) Metadata Debt:} The accumulated deficit of documentation
about data assets, their meaning, origin, ownership, quality
characteristics, and appropriate uses. Gartner predicted that 80\% of
data lakes would lack effective metadata management {[}10{]}, a concern
consistent with subsequent practitioner accounts. Prior work has shown that metadata quality, not merely its presence, is
what determines whether big data and unstructured content remain usable
\cite{r65}. Each day that data exists without proper metadata, the cost
of retrospective documentation increases as institutional knowledge
degrades. Internal APIs and data
pipelines frequently lack adequate documentation, creating continuity
risks that compound over time as personnel change.

\textbf{2) Quality Debt:} The accumulation of data quality issues,
duplicates, inconsistencies, missing values, outdated records, that were
not addressed at ingestion. Duplication in particular is a recurring and
tractable failure mode, and dedicated deduplication approaches have been
proposed to contain it at scale \cite{r66}. Industry research suggests that poor data
quality costs organizations millions annually, with Gartner estimating
average costs of \$12.9 million per organization {[}38{]}, {[}39{]}. The
compounding nature of quality debt is particularly severe: bad data
propagates through downstream systems, corrupting analytics, training
flawed machine learning models, and leading to cascading decision
errors. Industry surveys consistently report that data teams spend
substantial time on remediation rather than value creation.

\textbf{3) Security Debt:} The exposure created by deferred
implementation of access controls, encryption, and audit capabilities.
In the Data Lake context, the "store everything" mentality often meant
ingesting sensitive data without corresponding security controls,
whereas secure multi-tenant data services show that privacy controls can
be designed in from the start \cite{r67}.
Security debt carries catastrophic tail risk: a single breach can result
in regulatory fines, litigation, and reputational damage that dwarf the
cost of proactive security investment. According to enforcement tracking
databases, cumulative GDPR fines have exceeded \texteuro{}5 billion since 2018
{[}40{]}, with individual penalties reaching \texteuro{}1.2 billion for Meta
{[}41{]} and \texteuro{}746 million for Amazon {[}42{]}.

\textbf{4) Compliance Debt:} The gap between regulatory requirements and
actual data handling practices. Regulations like GDPR impose penalties
of up to \texteuro{}20 million or 4\% of annual global revenue {[}43{]}.
Organizations that built Data Lakes before GDPR (2018) often face
substantial compliance debt, data was collected without proper consent
mechanisms, stored without retention policies, and processed without the
documentation required for regulatory compliance. British Airways
received a \textsterling{}20 million fine from the UK ICO for data protection failures
(reduced from an initial \textsterling{}183 million notice of intention due to
COVID-19 impact considerations) {[}44{]}. Equifax\textquotesingle s 2017
breach resulted in settlements exceeding \$700 million with the FTC,
Consumer Financial Protection Bureau (CFPB), and state attorneys general
{[}45{]}.

\textbf{5) Organizational Debt:} The deficit of roles, processes, and
cultural norms required for effective data governance. This includes
missing data stewardship functions, undefined ownership, and
organizational resistance to governance discipline. Industry surveys
indicate that governance initiatives frequently fail when ownership
remains unclear. NewVantage Partners consistently finds that 92\% of
executives cite corporate culture as the primary barrier to becoming
data-driven {[}6{]}, a form of organizational debt that technology alone
cannot address.

\textbf{6) Operational Debt:} The deferred discipline of running the
platform once it is built. It accumulates when pipelines are not
supervised, when there is no observability, alerting, runbook, or
escalation path, when workloads are not isolated so a single job can
starve the cluster, and when capacity is planned without knowing the
workloads. Its root cause is the belief that delivery ends at go-live.
Operational debt is invisible on an architecture diagram, which is
exactly why it is under-reported, yet in the field catalogue it is among
the most frequent causes of a lake that technically works but is no
longer trusted or affordable to run.

\textbf{7) Engineering-Discipline Debt:} The deferred software
discipline of the data platform itself. It accumulates with the absence
of versioning for code, configurations, schemas, and datasets, the
absence of data CI/CD and of tests on transformations, missing
before-and-after reconciliation at migration, and hardcoded parameters,
paths, and credentials. The result is black-box pipelines and scripts
that no one can safely change, which freezes the platform and turns
every evolution into a risk. Like the others, this debt compounds: the
longer a pipeline runs untested and unversioned, the more expensive and
dangerous it becomes to touch.

\subsection{The Compound Interest Problem}

\begin{figure}[!t]
\centering
\includegraphics[width=\columnwidth]{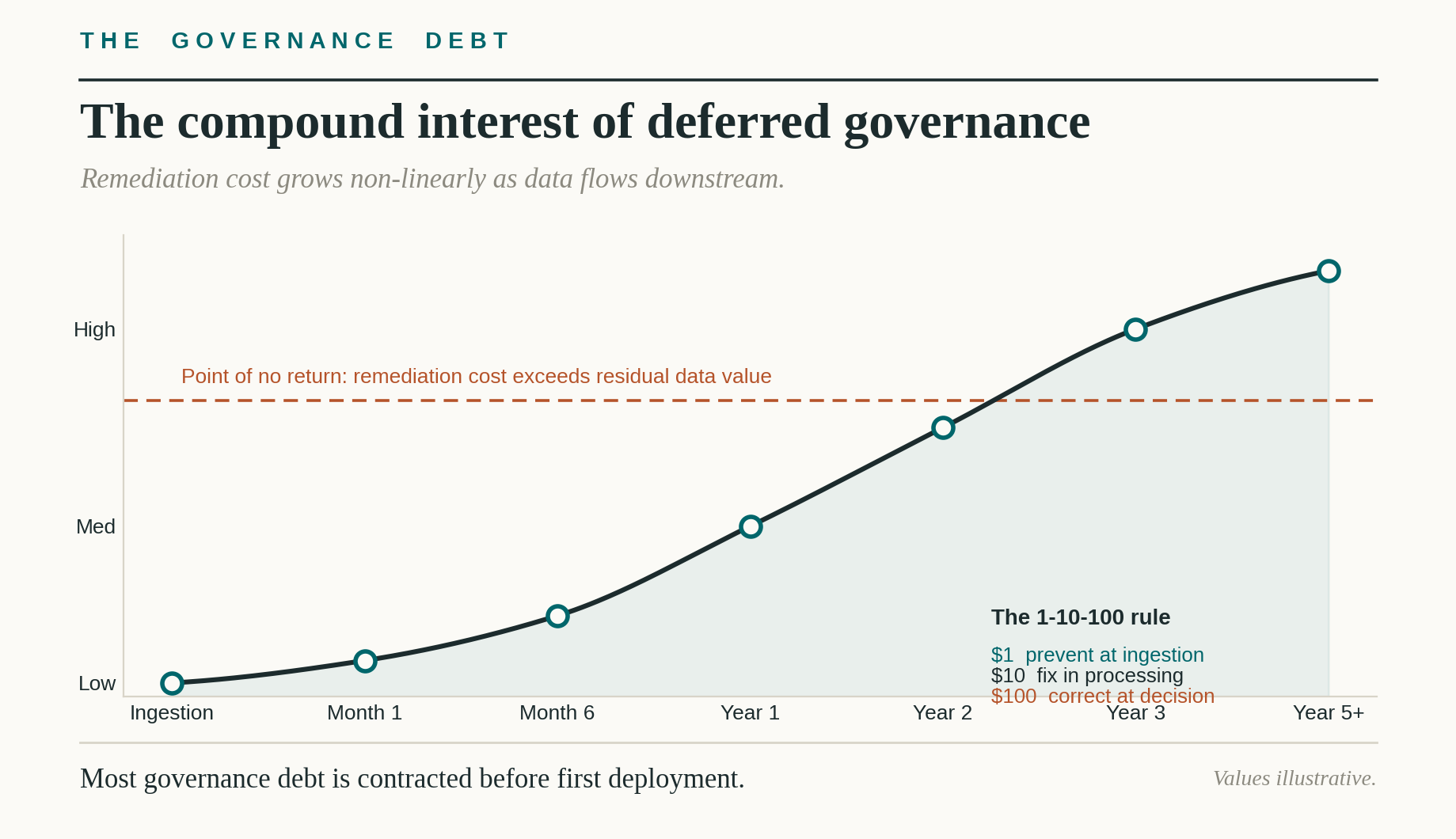}
\caption{The compound interest effect of governance debt. Remediation cost grows non-linearly as data flows downstream. Values are illustrative, reflecting the 1-10-100 heuristic widely cited in data quality practice.}
\label{fig:debt}
\end{figure}

What makes governance debt particularly dangerous is its compound
interest characteristics (Fig.~\ref{fig:debt}). Technical debt research demonstrates
that remediation costs grow significantly over time when issues remain
unaddressed, with delays substantially increasing eventual remediation
costs. Governance debt exhibits similar, and potentially more severe,
compounding effects, in at least four ways. The first is knowledge
degradation: the institutional understanding needed to interpret and
govern data erodes as people move on, so what could be documented in
hours at ingestion may take weeks of forensic work years later, if it
can be recovered at all, which is why early governance decisions, or
their absence, carry such outsized long-term weight. The second is
propagation: ungoverned data flows downstream into every derived
dataset, every pipeline, and every decision, and the 1-10-100 heuristic
common in data quality practice captures the escalation, roughly an
order of magnitude per stage from catching an issue at ingestion to
catching it at the point of decision; the exact ratios are illustrative
rather than validated, but the direction is consistently observed. The
third is the regulatory ratchet: rules tighten over time, so data
collected permissively in 2015 must now satisfy GDPR, the California
Consumer Privacy Act, and sector rules that did not exist at ingestion,
and debt incurred in a lenient regime comes due in a strict one, with
penalties attached. The fourth is volume amplification: as data grows
exponentially, debt that is manageable on a 10TB lake can become
insurmountable on a 10PB one, and McKinsey\textquotesingle s estimate
that 10 to 20\% of technology budgets are diverted to servicing
technical debt {[}35{]} is, if anything, conservative for governance
debt given its organizational reach.

\subsection{From Lake to Swamp: The Degradation Pathway}

The transformation from functional Data Lake to unusable Data Swamp
follows a predictable sequence that organizations rarely recognize until
remediation becomes prohibitively expensive. Understanding this
degradation pathway enables early intervention. Note that these stages
are not strictly sequential; they overlap and may run in parallel across
different organizational units or data domains. The timeline represents
typical progression in the absence of intervention:

\textbf{Stage 1, Rapid Ingestion (Months 1-6):} The lake launches
with emphasis on time-to-data. Multiple source systems connect rapidly.
Success is measured by ingestion volume and pipeline count. Governance
is acknowledged as important but deferred to "Phase 2" to avoid slowing
momentum. Initial users express enthusiasm.

\textbf{Stage 2, Governance Deferral (Months 6-18):} Data accumulates
faster than documentation. Metadata gaps emerge but are dismissed as
temporary. Schema variations proliferate as different teams ingest
similar data independently. Ownership remains undefined, everyone
assumes someone else is responsible. Early warning signs (duplicate
datasets, inconsistent naming, undocumented transformations) are visible
but not prioritized.

\textbf{Stage 3, Trust Erosion (Months 12-24):} Users encounter data
quality issues. Analysts discover conflicting values across tables.
Questions arise that nobody can answer: "Which customer table is
authoritative?" "Why do sales figures differ between reports?" "What
does this field actually mean?" Users develop workarounds, shadow
spreadsheets, private extracts, manual reconciliation. Lake usage
plateaus as trust declines.

\textbf{Stage 4, Retreat to Familiar Systems (Months 18-36):}
Business users return to the data warehouse for "trusted" data. Excel
spreadsheets proliferate for "flexibility." The lake becomes a staging
area rather than an analytical asset. IT maintains pipelines that nobody
uses. The original promise of democratization inverts: only specialists
can navigate the lake, and they prefer not to.

\textbf{Stage 5, Archive State (Months 24+):} The lake persists as
infrastructure but ceases to deliver value. Storage costs accumulate for
data that is neither used nor deleted (deletion requires governance
decisions that were never made). The lake becomes a liability, too
expensive to maintain, too risky to delete, too ungoverned to use.
Organizations begin planning "Data Lake 2.0" initiatives, often
repeating the pattern.

\textbf{Clarifying \textquotesingle Data Warehouse
Dependency\textquotesingle:} Before examining the forces that sustain
warehouse dependency, we must distinguish between two distinct meanings
of \textquotesingle data warehouse\textquotesingle{} that this analysis
treats differently. The \emph{data warehouse as trusted substrate}
refers to the institutional infrastructure that provides audit trails,
reconciliation controls, regulatory sign-off processes, and the
\textquotesingle official numbers\textquotesingle{} that finance,
compliance, and executives rely upon. This substrate function can be
implemented on traditional warehouse platforms (Teradata, Oracle, SQL
Server) or on modern cloud warehouses (Snowflake, BigQuery, Redshift),
but the distinguishing characteristic is not the technology, it is the
organizational investment in controls, ownership, and trust that the
infrastructure embodies. Separately, \emph{curated data layers} (data
marts, semantic layers, BI extracts, dimensional models) represent
structured, business-friendly views of data that may exist in
warehouses, lakehouses, or as virtualized layers. These curated layers
can technically be implemented anywhere. When we argue that
organizations \textquotesingle revert to the
warehouse,\textquotesingle{} we refer specifically to reversion to the
trusted substrate, the institutional controls and sign-off processes,
not merely to dimensional modeling or curated views. A lakehouse
implementing equivalent controls, ownership, and reconciliation would
constitute a new trusted substrate, not a reversion. The persistent
failures we document occur because organizations migrate data and
queries to lakes without migrating the institutional trust
infrastructure, then discover they cannot decommission the original
substrate because no equivalent exists.

\textbf{The Seven Forces of Data Warehouse Durability:} The retreat to
the data warehouse (Stage 4) reflects not merely behavioral inertia but
seven structural forces that make the warehouse inherently durable even
when lakes exist: (1) \emph{Regulatory reporting obligations}, auditors,
regulators, and boards require reproducible figures with complete audit
trails; the warehouse\textquotesingle s reconciliation controls and
sign-off processes satisfy these requirements while lakes typically
cannot. (2) \emph{Semantic layer and MDM heritage}, decades of
investment in conformed dimensions, business glossaries, and master data
hierarchies are encoded in warehouse structures; migrating these
requires not just technical effort but organizational consensus that
rarely materializes. (3) \emph{BI performance and query patterns},
warehouses are optimized for the star schemas and aggregation patterns
that dashboards and OLAP cubes require; lakes optimized for data science
workloads often underperform for traditional BI. (4) \emph{SQL-first
skills and tooling}, business analysts, finance teams, and most
enterprise users operate in SQL and Excel; lake technologies requiring
Python, Spark, or specialized notebooks create adoption barriers that
reinforce warehouse usage. (5) \emph{SLA contracts and data agreements},
formal service level agreements, data delivery schedules, and
inter-departmental contracts reference warehouse datasets; renegotiating
these agreements requires organizational effort that teams avoid. (6)
\emph{\textquotesingle Finance truth\textquotesingle{} anchoring}, the
CFO\textquotesingle s numbers, board presentations, and investor reports
derive from warehouse tables that have been reconciled to general ledger
for years; any alternative must demonstrate identical figures before
finance will trust it. (7) \emph{Usage migration cost}, migrating data
is straightforward; migrating hundreds of reports, dashboards, scheduled
jobs, and embedded analytics that reference warehouse tables requires
testing and validation effort that exceeds most project budgets. These
seven forces create structural lock-in that persists regardless of lake
maturity. Critically, these forces explain not merely Data Lake
\textquotesingle failure\textquotesingle{} but the observed pattern of
long-term coexistence: organizations maintain both warehouse and
lake/lakehouse infrastructures indefinitely because the forces prevent
complete warehouse decommission while lake investment continues for new
use cases. This coexistence, often lasting 5-10 years, represents a
stable equilibrium rather than a transitional state, with significant
cost and complexity implications that practitioners should anticipate.

\textbf{Governance Gravity, A Proposed Explanatory Mechanism:} Based on
patterns observed across the reviewed practitioner and analyst sources,
we hypothesize that these seven forces collectively manifest as what we
term \emph{governance gravity}, the pull of established governance
structures that organizations cannot escape without deliberately
replicating equivalent capabilities. We propose this concept as distinct
from \textquotesingle data gravity\textquotesingle, the physical
phenomenon where data\textquotesingle s mass attracts applications and
services due to latency and bandwidth costs of movement {[}46{]}. We
hypothesize that governance gravity operates at the organizational
layer: the pull derives not from data volume but from embedded semantic
contracts, audit trails, regulatory compliance, and institutionalized
trust that cannot be replicated by technical migration alone. The seven
durability forces cluster into three proposed mechanisms: (1)
institutional trust and auditability (forces 1, 6), the warehouse has
established controls, reconciliation processes, audit trails, and
accountable owners that the lake lacks, making it the natural refuge
when questions arise; (2) semantic anchoring (forces 2, 3), business
metrics and authoritative definitions are encoded in warehouse
structures (conformed dimensions, validated KPIs, certified reports)
that raw lake data cannot replicate without equivalent modeling effort;
and (3) operating model inertia (forces 4, 5, 7), teams, processes,
budgets, and SLAs are organized around the warehouse, relegating the
lake to experimental status until it can match these operational
guarantees. Expected observable signals of governance gravity strength
include: percentage of regulatory reports still sourced exclusively from
the warehouse (\textgreater80\% suggests strong gravity); proportion of
\textquotesingle gold\textquotesingle{} or
\textquotesingle certified\textquotesingle{} datasets residing only in
the warehouse versus the lake; and percentage of data quality rules
implemented in the warehouse but not replicated in the lake. These
thresholds (70\%, 80\%) are heuristics derived from practitioner
experience rather than empirically validated cutoffs; organizations
should calibrate them to their specific context. We propose the
following falsifiable prediction: when all three metrics exceed 70\%,
organizations will face structural lock-in that incremental lake
improvements cannot overcome. As reported across practitioner accounts,
organizations that attempt to decommission warehouses before
establishing equivalent governance in the lake discover these
dependencies through operational failure: regulatory reports that cannot
be reproduced, executive dashboards that diverge from audited figures,
and business definitions that nobody can authoritatively resolve.

\textbf{Conceptual Boundaries:} Governance gravity is distinct from
several related concepts in the literature. Unlike \emph{data gravity}
(physical: latency and bandwidth costs of moving data), governance
gravity operates at the organizational layer through semantic contracts
and institutional trust. Unlike \emph{technical debt} (accumulated code
shortcuts requiring future rework), governance gravity describes
structural dependencies that may be well-engineered rather than
shortcuts, the warehouse\textquotesingle s governance is not
\textquotesingle debt\textquotesingle{} but rather an asset the lake
fails to replicate. Unlike \emph{organizational inertia} (general
resistance to change), governance gravity identifies specific,
measurable dependencies that rational actors maintain because
alternatives do not yet provide equivalent guarantees. Unlike
\emph{vendor lock-in} (switching costs created by proprietary
technologies), governance gravity persists even when technologies are
open-source and portable, the lock-in derives from organizational
knowledge and process embedding, not vendor contracts.

Governance gravity can in principle be measured. We sketch a set of
candidate parity-gap metrics between warehouse and lake, together with a
protocol for validating them, but because they rest on unvalidated
numeric thresholds we keep them out of the main argument and collect them
in Appendix~\ref{app:numeric}.

\textbf{A Recurring Field Case:} The following is a composite of
situations I have repeatedly encountered, anonymized to respect client
confidentiality rather than invented for illustration. A bank migrates
its risk reporting to the Data Lake. Months later, regulators request a
historical trade reconciliation, and the lake cannot reproduce figures
matching the audited warehouse reports: timestamp granularity differs
between systems, currency conversion is applied at different points in
the pipeline, and the definition of ``trade date'' varies across source
systems. Leadership faces a choice between rebuilding the entire pipeline
with warehouse-equivalent controls, a multi-quarter effort, and reverting
to the warehouse for regulatory reporting while the lake serves only
exploratory analytics. In the cases I observed, the organization reverted:
a textbook manifestation of governance gravity. The lake had reached
technical parity but not governance parity. It could store and query the
same data, yet it could not provide the semantic consistency, audit
trail, and reconciliation guarantees that regulatory compliance demanded.

\textbf{The Semantic Layer Capability Gap:} This case illustrates a
fundamental capability gap: Data Lakes replace storage and query
infrastructure but not the semantic layer that warehouses provide. The
semantic layer encompasses business glossaries with authoritative term
definitions, certified metrics with documented calculation logic,
conformed dimensions ensuring consistent entity representation across
reports, and reconciliation rules that guarantee financial figures tie
out. Without explicit reconstruction of this layer, organizations
effectively operate two parallel truth systems, the
warehouse\textquotesingle s certified semantics versus the
lake\textquotesingle s raw data, creating confusion, duplication of
effort, and inevitable regression to the trusted system. Table VI maps
the governance artifacts that must be replicated.

\begin{table*}[!t]
\centering
\footnotesize
\renewcommand{\arraystretch}{1.25}
\caption{DATA WAREHOUSE GOVERNANCE ARTIFACTS AND LAKE EQUIVALENTS}
\begin{tabular}{@{}@{}
  >{\raggedright\arraybackslash}p{(\columnwidth - 6\tabcolsep) * \real{0.2295}}
  >{\raggedright\arraybackslash}p{(\columnwidth - 6\tabcolsep) * \real{0.2370}}
  >{\raggedright\arraybackslash}p{(\columnwidth - 6\tabcolsep) * \real{0.2693}}
  >{\raggedright\arraybackslash}p{(\columnwidth - 6\tabcolsep) * \real{0.2641}}@{}}
\toprule
\textbf{DW Artifact} & \textbf{Lake Equivalent} & \textbf{Risk if
Absent} & \textbf{Typical Tools} \\
\midrule
Conformed dimensions & Centralized dimension tables with CDC & Entity
inconsistency across reports & dbt, Spark dimension processing \\
Business glossary / MDM & Data catalog with business terms & Semantic
ambiguity, multiple truths & Collibra, Alation, Atlan, Datahub \\
Certified metrics / KPIs & Metrics layer with versioning & KPI
divergence, trust erosion & dbt metrics, Cube, Looker, AtScale \\
Reconciliation controls & Automated tie-out jobs & Undetected drift,
audit failures & Great Expectations, Soda, dbt tests \\
Audit trail / lineage & Column-level lineage tracking & Regulatory
non-compliance & OpenLineage, Marquez, Unity Catalog \\
Change control process & Data contracts + CI/CD & Breaking changes,
downstream failures & Schemata, Buf, Protobuf, JSON Schema \\
SLA guarantees & Freshness + quality SLOs & Stale data, missed
commitments & Monte Carlo, Bigeye, Anomalo \\
Named data stewardship & Domain ownership registry & Accountability
void, no escalation & Catalog ownership fields, RACI \\
\bottomrule
\end{tabular}
\end{table*}

\textbf{Reconciliation as Non-Negotiable:} In mature data warehouses,
reconciliation controls are institutionalized: balance checks verify
that debits equal credits, tie-outs confirm that subsidiary ledgers sum
to general ledger totals, and periodic reconciliations catch drift
before it compounds. These controls create the \textquotesingle trust
infrastructure\textquotesingle{} that enables regulatory sign-off and
executive confidence. Data Lakes that fail to institutionalize
equivalent controls, automated variance detection, cross-system
tie-outs, and alerting on anomalies, cannot achieve the trust threshold
required for mission-critical workloads. The absence of reconciliation
creates a \textquotesingle trust gap\textquotesingle{} that typically
drives users back to Excel or the warehouse, regardless of the
lake\textquotesingle s technical capabilities.

\textbf{Achieving Governance Parity:} To neutralize governance gravity,
organizations must systematically reconstruct the governance artifacts
that warehouses provide implicitly. Required artifacts include: (1) a
certified metrics repository with validated KPI definitions, calculation
logic, and version history; (2) conformed dimensions with authoritative
master data and change tracking; (3) automated reconciliation controls
that detect drift between systems and alert on variance; (4)
comprehensive audit trails meeting regulatory retention requirements;
(5) certified datasets with documented quality guarantees, freshness
SLAs, and named owners; and (6) explicit ownership assignments with
escalation paths for disputes. Migration proceeds through dual-run
phases where regulatory and audit-critical workloads execute in both
systems until reconciliation validates equivalence over a defined
period. Cutover criteria should specify acceptable variance thresholds
(typically zero for financial figures), required reconciliation duration
(often 2-3 reporting cycles), and rollback triggers. This systematic
approach transforms governance gravity from an insurmountable obstacle
into a structured checklist, demanding but achievable.

This degradation pathway is not inevitable but is remarkably common. The
key intervention point is Stage 2, where governance debt begins
accumulating but remediation remains feasible. By Stage 4, remediation
costs typically exceed the value of the existing data, making
replacement more economical than repair. Organizations that recognize
swamp indicators early, using criteria such as those in the operational
definition above, can arrest degradation before it becomes irreversible.

\textbf{The Legacy Amplifier:} Legacy systems do not merely provide
context for swamp formation, they actively drive it through three
mechanisms. First, \emph{legacy constraints}: batch windows that prevent
real-time synchronization, incomplete change data capture that misses
critical updates, upstream quality issues that cannot be corrected at
source, and proprietary formats that resist standardization. Second,
\emph{contract rigidity}: schemas frozen by decades of downstream
dependencies, where any evolution requires coordinated changes across
multiple systems that rarely align on timelines or priorities. Third,
\emph{ownership fragmentation}: legacy systems belong to business units
or IT domains separate from the lake team, creating accountability gaps
where neither party owns data quality end-to-end. These factors mean
that swamp formation is not solely attributable to \textquotesingle poor
governance\textquotesingle{} but often reflects governance that is
operationally infeasible at scale without first addressing legacy
dependencies that predate the lake initiative.

\textbf{Legacy Constraint Categories:} These mechanisms cluster into
four constraint categories that systematically feed the seven durability
forces identified above: (1) \emph{Technical constraints} (batch
windows, CDC limitations, performance bottlenecks, proprietary formats)
directly feed durability forces 3 and 4 (BI performance expectations and
SQL-first skills), the lake cannot match warehouse response times for
legacy query patterns, reinforcing tool preferences. (2)
\emph{Regulatory constraints} (audit trail requirements, retention
policies, sign-off processes, evidence preservation) directly feed
durability forces 1 and 6 (regulatory reporting and finance truth
anchoring), compliance obligations cannot be satisfied by infrastructure
lacking equivalent controls. (3) \emph{Contractual constraints} (frozen
interfaces, vendor lock-in, SLA commitments, integration agreements)
directly feed durability force 5 (SLA contracts), formal agreements
reference warehouse datasets and cannot be unilaterally renegotiated.
(4) \emph{Organizational constraints} (SQL-dominant skills, historical
ownership patterns, embedded processes, tribal knowledge) directly feed
durability forces 2, 4, and 7 (semantic heritage, SQL-first skills,
usage migration cost), organizational capability is structured around
warehouse paradigms that lakes do not replicate. This mapping
demonstrates that governance gravity is not a single phenomenon but
rather the emergent result of multiple constraint categories reinforcing
multiple durability forces simultaneously, explaining why addressing any
single constraint rarely reduces overall gravity.

\textbf{Legacy Archetypes:} These mechanisms manifest differently across
four common legacy archetypes that practitioners will recognize. (1)
\emph{Monolithic ERP/CRM systems} (SAP, Oracle, Salesforce legacy
instances): rigid schemas with thousands of custom fields accumulated
over decades; upgrade cycles that span years and freeze extraction
interfaces; business logic embedded in application layer that cannot be
replicated in the lake. (2) \emph{Mainframe and batch processing
systems}: nightly batch windows that create 24-hour latency floors;
COBOL copybooks and EBCDIC encoding that require specialized
transformation; scheduled job dependencies that cannot tolerate
disruption to established sequences. (3) \emph{SQL/PLSQL procedural
heritage}: thousands of stored procedures encoding business rules that
exist nowhere else; \textquotesingle tribal knowledge\textquotesingle{}
where only one or two individuals understand critical transformations;
procedures that have evolved over 15-20 years and would require months
to reverse-engineer. (4) \emph{Legacy BI semantic layers} (Business
Objects universes, Cognos Framework Manager models, MicroStrategy
schemas): years of metric definitions, calculated fields, and security
rules that are not portable to modern tools; hundreds of reports built
against these layers that would require migration if the semantic layer
changes. Each archetype creates distinct constraints that shape how
governance debt accumulates and why remediation proves difficult.

\textbf{Legacy-to-Swamp Causal Pathways:} The legacy constraint
categories feed directly into specific swamp indicators through
observable causal pathways: (1) \emph{Technical constraints $\rightarrow$ Lineage
gaps and metadata incompleteness.} Batch windows and partial CDC prevent
real-time lineage tracking; proprietary formats and COBOL copybooks
resist automated metadata extraction; transformation logic embedded in
legacy scripts cannot be parsed by modern lineage tools. Observable
signal: lineage coverage drops sharply at legacy system boundaries. (2)
\emph{Organizational constraints $\rightarrow$ Ownership absence and unused
datasets.} Fragmented ownership across legacy system boundaries means no
single team owns end-to-end data quality; tribal knowledge concentrated
in legacy specialists who do not participate in lake governance;
datasets ingested \textquotesingle because the source
exists\textquotesingle{} rather than for articulated use cases remain
orphaned. Observable signal: ownership coverage inversely correlates
with number of legacy source systems. (3) \emph{Regulatory constraints $\rightarrow$
Quality validation gaps and duplication.} Audit requirements force
retention of legacy validation logic in original systems rather than
migration to lake; parallel validation creates duplicate datasets
maintained for compliance rather than consolidated; sign-off processes
anchored to legacy systems prevent quality gate migration. Observable
signal: validation coverage highest for born-in-lake datasets, lowest
for legacy-sourced data. (4) \emph{Contractual constraints $\rightarrow$
Documentation debt.} Frozen interfaces documented decades ago with no
maintained data dictionary; SLA commitments referencing field names
without semantic definitions; vendor contracts specifying data formats
without business context. Observable signal: metadata completeness
declines with dataset age and number of integration hops from source.
This mapping transforms \textquotesingle legacy\textquotesingle{} from a
narrative explanation into a set of testable hypotheses about swamp
indicator drivers.

\textbf{Legacy Debt Taxonomy:} The constraints imposed by legacy systems
accumulate as four distinct categories of \textquotesingle legacy
debt\textquotesingle{} that interact with and amplify governance debt
(Table VII).

\begin{table*}[!t]
\centering
\footnotesize
\renewcommand{\arraystretch}{1.25}
\caption{LEGACY DEBT TAXONOMY}
\begin{tabular}{@{}@{}
  >{\raggedright\arraybackslash}p{(\columnwidth - 6\tabcolsep) * \real{0.2217}}
  >{\raggedright\arraybackslash}p{(\columnwidth - 6\tabcolsep) * \real{0.2898}}
  >{\raggedright\arraybackslash}p{(\columnwidth - 6\tabcolsep) * \real{0.2316}}
  >{\raggedright\arraybackslash}p{(\columnwidth - 6\tabcolsep) * \real{0.2569}}@{}}
\toprule
\textbf{Debt Category} & \textbf{Manifestation} & \textbf{Impact on
Lake} & \textbf{Remediation Approach} \\
\midrule
\textbf{Data Model Debt} & Frozen schemas with implicit semantics;
undocumented field meanings; accumulated custom fields & Semantic
ambiguity propagates; interpretation varies by analyst & Schema
documentation sprint; field-level business glossary \\
\textbf{Pipeline Debt} & Batch windows; partial CDC; undocumented
extract scripts; tribal knowledge transformations & Latency floors; data
gaps; transformation opacity & CDC modernization; pipeline
documentation; script reverse-engineering \\
\textbf{Organizational Debt} & Fragmented ownership;
\textquotesingle who owns quality?\textquotesingle{} ambiguity; siloed
accountability & No end-to-end quality owner; issues fall between teams
& Domain ownership model; RACI clarification; stewardship program \\
\textbf{Compliance Debt} & Audit/traceability externalized to DW;
regulatory reporting locked to warehouse & Lake cannot satisfy
regulatory requirements; reversion mandatory & Lineage implementation;
audit trail replication; dual-run validation \\
\bottomrule
\end{tabular}
\end{table*}

These four debt categories compound: data model debt creates semantic
ambiguity that pipeline debt propagates, organizational debt ensures no
one addresses the accumulating issues, and compliance debt ultimately
forces reversion when regulators demand auditability the lake cannot
provide. Successful lake initiatives must address all four categories,
not merely technical integration.

\textbf{The Legacy Reinforcement Loop:} These mechanisms create a
self-reinforcing cycle that entraps organizations: legacy constraints
force ingestion compromises $\rightarrow$ compromises accumulate as governance debt
$\rightarrow$ debt erodes trust in lake data $\rightarrow$ eroded trust drives shadow pipelines
and workarounds $\rightarrow$ shadow pipelines increase complexity $\rightarrow$ complexity
makes legacy decommissioning impossible $\rightarrow$ legacy remains entrenched $\rightarrow$
constraints persist. This loop explains the non-linear, compound
interest nature of governance debt (Table IX): each iteration through
the cycle amplifies the debt burden, as knowledge loss, propagation
effects, and complexity growth compound upon each preceding cycle.
Breaking the loop requires explicit intervention at multiple points
simultaneously: addressing source quality, establishing governance,
building trust, and managing complexity, a coordination challenge that
explains why incremental improvement strategies routinely fail.

\textbf{Legacy Exit Patterns:} While legacy dependencies constrain
governance, established patterns exist for managing the transition
{[}47{]}, {[}48{]}, {[}49{]}. The \emph{Strangler Fig} pattern {[}47{]}
incrementally replaces legacy components by routing new functionality
through the lake while maintaining legacy interfaces, suitable for
modular systems with extended timelines; primary risk is interface
governance complexity as the number of integration points proliferates.
\emph{CDC with Dual-Run} {[}49{]} employs Change Data Capture with
parallel operation in both systems, enabling validation before cutover,
essential when continuity is non-negotiable; primary risks are drift
between systems and ongoing reconciliation cost during extended dual-run
periods. \emph{Contractual Freeze with Façade} freezes legacy schemas
while exposing data through API layers, appropriate when legacy systems
cannot be modified; primary risk is that the façade becomes permanent
technical debt rather than a transitional measure. \emph{Data Contracts}
{[}48{]} establish explicit agreements between producers and consumers
on schema, quality, and SLAs, addressing ownership fragmentation
directly; primary risks are adoption resistance from data producers and
enforcement gaps when contracts are violated. Finally, \emph{Source
Remediation} invests in correcting quality issues at origin rather than
compensating downstream, offering highest long-term value; primary risk
is that upstream ownership cooperation is rarely available, making this
pattern aspirational for many organizations. The appropriate pattern
depends on legacy system characteristics, organizational constraints,
and acceptable transition timelines. As guidance: Strangler Fig and CDC
with Dual-Run primarily address legacy constraints (batch windows,
incomplete CDC); Contractual Freeze addresses contract rigidity (frozen
schemas); Data Contracts address ownership fragmentation (accountability
gaps); Source Remediation addresses all three but requires the highest
organizational alignment.

The following mapping demonstrates how degradation stages, swamp
indicators, and GDAM dimensions form an integrated diagnostic system
(Table VIII):

\begin{table*}[!t]
\centering
\footnotesize
\renewcommand{\arraystretch}{1.25}
\caption{INTEGRATED DIAGNOSTIC MAPPING: STAGES, SWAMP INDICATORS, AND GDAM}
\begin{tabular}{@{}@{}
  >{\raggedright\arraybackslash}p{(\columnwidth - 4\tabcolsep) * \real{0.2786}}
  >{\raggedright\arraybackslash}p{(\columnwidth - 4\tabcolsep) * \real{0.4819}}
  >{\raggedright\arraybackslash}p{(\columnwidth - 4\tabcolsep) * \real{0.2395}}@{}}
\toprule
\textbf{Stage} & \textbf{Primary Swamp Indicators} & \textbf{GDAM
Dimensions} \\
\midrule
Stage 2: Governance Deferral & Ownership gaps, Metadata incomplete,
Lineage undocumented & MC, OO, LT \\
Stage 3: Trust Erosion & Unused datasets, Quality issues, Conflicting
values & QO, LT \\
Stage 4: Retreat to DW & All indicators compound; governance gravity
activates & AG, OO \\
Stage 5: Archive State & Duplication endemic, No validation, Full
dysfunction & All (debt crystallized) \\
\bottomrule
\end{tabular}
\end{table*}

This mapping demonstrates that the degradation pathway, diagnostic
indicators, and maturity assessment form an integrated system:
early-stage interventions targeting specific GDAM dimensions
(particularly MC and OO in Stage 2) can arrest progression before
governance gravity takes hold in Stage 4. Recommended priority actions
by stage: Stage 2: establish ownership assignments and metadata
requirements; Stage 3: implement data quality gates and certification
processes; Stage 4: achieve governance parity with reconciliation
automation before attempting DW decommission; Stage 5: evaluate rebuild
versus remediation based on residual value assessment.

\subsection{The Breaking Point: When Governance Debt Becomes Unsustainable}

Research on technical debt suggests that software organizations reach
unsustainable levels when maintenance activities consume the majority of
development capacity {[}36{]}. While specific thresholds vary by
context, the general pattern, that accumulated maintenance burden
eventually crowds out value creation, applies to governance debt.

We observe three breaking points in Data Lake governance, and they tend
to arrive in sequence. The first is a collapse of trust. Once quality
problems become pervasive enough that users stop believing the lake's
contents, the asset turns into a liability: analytics teams build shadow
systems, business users drift back to spreadsheets, and the platform
persists while serving no one. This is the state Gartner named a data
swamp in 2014 {[}3{]}. The second arrives with a compliance shock. A
regulatory inquiry, an audit, or a breach exposes how much compliance
debt has accrued, and the organization finds it cannot say what data it
holds, where it came from, who has touched it, or whether consent was
ever obtained. Capital One's 2019 breach, which exposed more than a
hundred million customer records and drew an eighty-million-dollar fine
from the Office of the Comptroller of the Currency {[}50{]}, is the kind
of event that makes that debt suddenly visible. The third is
remediation impossibility. Past a certain point, retrospective governance
simply costs more than rebuilding from scratch, sometimes more than the
data is worth at all, and the organization abandons the lake and starts
again, having paid an expensive tuition in governance debt.

\subsection{Implications for Practice}

The governance debt framework has several practical implications. First,
governance is not optional overhead but a form of investment that
prevents future liabilities. Organizations should budget for governance
as they budget for security, not as discretionary spending but as risk
management. Second, early governance decisions have outsized impact. The
compound interest nature of governance debt means that decisions made
(or deferred) in the first months of a Data Lake initiative have
consequences that persist for years. Third, governance debt should be
made visible. Like technical debt, governance debt exists whether or not
it appears on balance sheets. Organizations benefit from explicit
accounting of governance debt to inform strategic decisions about
remediation versus replacement.

\subsection{Governance Debt Assessment Model (GDAM)}

To operationalize the governance debt concept for practitioners, we
propose the Governance Debt Assessment Model (GDAM): a qualitative
rubric across five dimensions, not a numeric score. We deliberately
avoid a 0 to 20 scale and fixed percentage thresholds, which would
manufacture a precision the instrument does not have. Each dimension is
instead read at one of four qualitative levels, \emph{Absent}, \emph{Ad
hoc}, \emph{Partial}, or \emph{Established}, and the diagnostic value
lies in the profile across the five dimensions rather than in any
aggregate number.

\textbf{Dimension 1, Metadata Completeness:} how widely data assets
carry complete business metadata (definition, owner, lineage), ranging
from Absent, where critical assets are undocumented, to Established,
where documentation is the default condition of ingestion.

\textbf{Dimension 2, Quality Observability:} the reach of automated
quality monitoring, from Absent (no monitoring), through Ad hoc checks
and coverage of critical pipelines only, to Established (comprehensive
monitoring with alerting and SLAs).

\textbf{Dimension 3, Access Governance:} the coverage of policy
enforcement, from Absent or basic authentication, through role-based
access on some assets, to Established attribute-based access with audit
trails and automated compliance.

\textbf{Dimension 4, Lineage Traceability:} the reach of end-to-end
lineage, from Absent or purely manual, through partial automation, to
Established automated lineage with impact analysis.

\textbf{Dimension 5, Organizational Ownership:} the extent of effective
data stewardship, from Absent or nominal, through ownership for a
minority of assets, to Established accountability with active
stewardship.

Read together, the five levels give a profile rather than a grade. An
organization that is Absent or Ad hoc on most dimensions sits in the
zone where remediation costs climb steeply if debt keeps accumulating,
and the rubric points to the weakest dimensions as the first targets.
This is a structured prompt for diagnosis, offered for practitioners to
adapt, and it awaits empirical validation. Potential validation
approaches include: (1) multi-case study design comparing
GDAM profiles across organizations with documented lake outcomes (success,
swamp, abandonment) to establish criterion validity; (2) inter-rater
reliability assessment using independent evaluators scoring the same
organizations to establish measurement consistency; (3) correlation
analysis between governance profiles and operational metrics such as
data request lead time, data quality incident frequency, and user
adoption rates; and (4) longitudinal tracking to test whether weak
profiles predict subsequent governance crises, establishing predictive
validity. Such validation would strengthen the
framework\textquotesingle s diagnostic and predictive utility while
enabling refinement of the dimensions and their levels. We keep the GDAM
to its five governance dimensions on purpose. The two
components added in this paper, operational debt and
engineering-discipline debt, are diagnosed qualitatively through the
field catalogue rather than folded into the rubric: adding two further
scales would reintroduce the false precision we are avoiding. A reader
should therefore treat the GDAM profile as a structured prompt for the
five classic dimensions, and the catalogue clusters of Section VIII as
the diagnostic for the operational and engineering dimensions.

\textbf{Preliminary Expert Review:} To assess face validity and
plausibility prior to formal empirical
validation, the framework components (GDAM dimensions, swamp indicators,
governance gravity reasoning) were reviewed by eight
practitioners with direct Data Lake implementation experience: three
enterprise data architects (10+ years), two data governance program
managers, one chief data officer, and two consultants specializing in
data platform migrations. Reviewers were asked to evaluate: (1) whether
the dimensions and indicators align with their observed failure
patterns; (2) whether the proposed thresholds appear plausible given
their experience; and (3) what refinements they would suggest. All eight
confirmed that the GDAM dimensions capture the primary governance
failure modes they have observed. They agreed that a profile sitting at
\emph{Absent} or \emph{Ad hoc} across most dimensions corresponds to the
High-Risk zone they recognize, while several cautioned that the exact
tipping point varies by organization. Reviewers endorsed the directional
logic of the swamp indicators and of governance gravity, while noting
that any specific cut-off would need calibration and likely varies by
industry regulatory intensity. This preliminary review establishes face
validity but does not constitute empirical validation; the rubric and
indicators should be treated as testable hypotheses rather than
prescriptive standards.

\textbf{Expert Panel Protocol:} Reviewers were recruited through
professional networks and selected for diversity of role, sector, and
geography. Each reviewer received a briefing document summarizing the
framework components and proposed thresholds. Semi-structured interviews
(45-60 minutes each) explored threshold plausibility, alignment with
observed patterns, and suggested refinements. Responses were
consolidated using thematic analysis to identify areas of consensus and
divergence. Panel composition: three enterprise data architects
(financial services, healthcare, retail; 10-15 years experience; North
America and Europe), two data governance program managers (insurance,
manufacturing; 8-12 years; North America), one chief data officer
(technology sector; 20 years; Europe), and two consultants (Big Four and
boutique; 12-18 years; global). This diversity ensures the framework was
evaluated against varied organizational contexts, though the sample size
precludes statistical generalization.

\textbf{Worked Example (illustrative):} To show how the instrument is
read, rather than to assert measured values, consider an organization
three years into a Data Lake initiative whose catalogue profile is
typical of a struggling program: metadata documented only for critical
datasets, lineage maintained manually and only for regulatory reports,
ownership assigned for a minority of datasets, monitoring confined to
production pipelines, and role-based access in place but with incomplete
audit trails. A reviewer applying the GDAM to this profile would score
Metadata Completeness, Lineage Traceability, and Organizational
Ownership low, Quality Observability and Access Governance in the middle
band, and would land in the High-Risk classification. The instrument's
value is not the aggregate number but the direction it gives: it points
to ownership and metadata as the quickest-return interventions and flags
that remediation should begin before Stage 4, where governance gravity
activates. The inputs above are illustrative placeholders chosen to
exercise the scoring logic, not observed measurements, consistent with
the qualitative stance of this paper.

\textbf{Relationship to Swamp Indicators:} The swamp indicators
introduced in Section V serve as rapid diagnostic signals, while GDAM
provides a more granular maturity assessment. The two instruments are
complementary: swamp indicators enable quick detection of degradation;
GDAM enables root cause analysis and remediation planning. The mapping
between them is as follows: absence of ownership maps to Organizational
Ownership (OO); metadata incompleteness maps to Metadata Completeness
(MC); undocumented lineage maps to Lineage Traceability (LT); excessive
duplication and missing quality validation both map to Quality
Observability (QO); and access/compliance gaps map to Access Governance
(AG). Unused datasets represent a symptom that may stem from failures in
multiple dimensions (typically MC and OO). An organization whose GDAM
profile sits mostly at Absent or Ad hoc will typically exhibit
three or more swamp indicators, confirming that the rapid diagnostic and
the detailed assessment converge on the same underlying dysfunction.

\textbf{The Causal Chain, Swamp to Reversion (Proposed Mechanism):}
Based on patterns observed across the reviewed practitioner and analyst
sources, we propose that the relationship between swamp formation,
governance gravity, and warehouse reversion follows a predictable causal
sequence. We hypothesize four stages with expected observable signals:
\emph{Stage 1: Swamp formation erodes trust.} As swamp indicators
accumulate (absent ownership, incomplete metadata, undocumented
lineage), users lose confidence in lake data quality and reliability.
Expected observable signals: Finance cannot reconcile figures; analysts
produce conflicting reports; executives receive inconsistent dashboards.
\emph{Stage 2: Trust erosion widens the governance parity gap.} The
parity gap metrics (KPI divergence, reconciliation volume, dual-run
duration) deteriorate as swamp conditions persist. Expected observable
signals: Lake-sourced KPIs diverge from warehouse
\textquotesingle source of truth\textquotesingle; manual reconciliation
hours increase as teams validate lake outputs against trusted systems;
dual-run periods extend indefinitely because no one will sign off on
warehouse decommission. \emph{Stage 3: Widening parity gap activates
governance gravity.} The seven durability forces (regulatory
obligations, semantic heritage, SLA contracts, etc.) become binding
constraints rather than theoretical concerns. Expected observable
signals: regulatory reports cannot be reproduced from lake data; CFO
dashboards diverge from audited figures; SLA commitments are missed, the
forces pull organizations back to infrastructure that provides required
guarantees. \emph{Stage 4: Gravity triggers reversion behaviors.} As
reported across Tier-2 and Tier-3 evidence, observable reversion events
include: creation of shadow data marts sourced from warehouse rather
than lake; migration of regulatory and executive reports back to
warehouse infrastructure; expansion of warehouse capacity despite stated
\textquotesingle lake-first\textquotesingle{} strategy; proliferation of
Excel extracts bypassing the lake entirely; and formal deprioritization
of lake investment in favor of
\textquotesingle trusted\textquotesingle{} platforms. We propose the
following falsifiable prediction: swamp remediation, if delayed past
Stage 2, will prove insufficient because by Stage 3-4, governance
gravity has activated forces that cannot be neutralized by improving
lake data quality alone.

We also explored a more speculative, numeric layer: a decision rule
linking sustained swamp indicators to reversion, and a swamp severity
score meant to flag organizations heading for it. Both depend on
thresholds we cannot yet validate, so they live in
Appendix~\ref{app:numeric} rather than in the main argument.

\begin{table*}[!t]
\centering
\footnotesize
\renewcommand{\arraystretch}{1.25}
\caption{GOVERNANCE DEBT COMPONENTS AND ASSOCIATED COSTS}
\begin{tabular}{@{}@{}
  >{\raggedright\arraybackslash}p{(\columnwidth - 6\tabcolsep) * \real{0.2513}}
  >{\raggedright\arraybackslash}p{(\columnwidth - 6\tabcolsep) * \real{0.3016}}
  >{\raggedright\arraybackslash}p{(\columnwidth - 6\tabcolsep) * \real{0.2658}}
  >{\raggedright\arraybackslash}p{(\columnwidth - 6\tabcolsep) * \real{0.1814}}@{}}
\toprule
\textbf{Debt Type} & \textbf{Manifestation} & \textbf{Documented Cost} &
\textbf{Source} \\
\midrule
Metadata Debt & Missing documentation, unknown lineage, degraded
knowledge & 80\% lack metadata management & Gartner {[}10{]} \\
Quality Debt & Duplicates, inconsistencies, missing values, stale data &
Avg. \$12.9M annual cost (Gartner est.) & Gartner {[}38{]} \\
Security Debt & Inadequate access controls, missing encryption, exposure
risk & \texteuro{}5.88B cumulative GDPR fines & CMS Tracker {[}40{]} \\
Compliance Debt & Gap between regulatory requirements and actual
practices & Up to \texteuro{}20M or 4\% global revenue & EU GDPR {[}43{]} \\
Organizational Debt & Missing roles, unclear ownership, cultural
resistance & 92\% cite culture as primary barrier & NVP Survey
{[}6{]} \\
Operational Debt & No supervision, observability, runbooks, or workload
isolation; the run unfunded & Lake works but is untrusted and costly to
operate & Field catalogue (this paper) \\
Engineering-Discipline Debt & No versioning, CI/CD, or tests; hardcoded
secrets; black-box pipelines & Platform frozen; every change becomes a
risk & Field catalogue (this paper) \\
\bottomrule
\end{tabular}
\end{table*}

\section{DID WE LEARN? TECHNICAL PROGRESS AND PERSISTENT CHALLENGES}

Failure has a way of breeding successors. The data lake's troubles gave
us the Data Lakehouse, Data Mesh, Data Fabric, and lately the AI-ready
platform, each pitched as the fix for the one before it (Table X). The
fair question is not whether the technology improved. It plainly did. The
question is whether the organizational failures went with it.

\subsection{Genuine Technical Improvements}

Successor paradigms have delivered substantive technical advances that
address legitimate limitations of original Data Lake architectures.
These improvements merit acknowledgment:

\textbf{ACID Transaction Support:} Open table formats (Delta Lake,
Apache Iceberg, Apache Hudi) provide transactional guarantees previously
unavailable in distributed file systems. This enables reliable
concurrent writes, schema evolution, and time-travel queries,
capabilities that eliminate entire categories of data integrity issues.

\textbf{Unified Batch and Streaming:} Modern architectures support both
batch and streaming workloads on unified infrastructure, reducing
architectural complexity and enabling near-real-time analytics without
maintaining separate systems.

\textbf{Native Governance Tooling:} Lakehouse platforms increasingly
embed data cataloging, lineage tracking, and access control as
first-class features rather than afterthoughts. Unity Catalog, AWS Glue
Data Catalog, and similar services provide out-of-the-box governance
capabilities.

\textbf{Data Contracts and Quality Gates:} Data Mesh principles
formalized the concept of data contracts, explicit agreements about
schema, quality, and service levels between producers and consumers.
Tools like Great Expectations and dbt tests enable enforcement of
quality constraints at pipeline boundaries.

\textbf{SQL Accessibility:} Direct SQL querying of lake data eliminates
the need for data movement to separate warehouses, reducing latency,
cost, and governance complexity while enabling broader access by
analysts familiar with SQL but not distributed computing frameworks.

These technical advances are genuine and valuable. However, they address
symptoms rather than root causes. Table X summarizes what successor
paradigms genuinely resolve versus what remains unaddressed.

\begin{table*}[!t]
\centering
\footnotesize
\renewcommand{\arraystretch}{1.25}
\caption{SUCCESSOR PARADIGMS: GENUINE FIXES VS. PERSISTENT GAPS}
\begin{tabular}{@{}@{}
  >{\raggedright\arraybackslash}p{(\columnwidth - 4\tabcolsep) * \real{0.2418}}
  >{\raggedright\arraybackslash}p{(\columnwidth - 4\tabcolsep) * \real{0.3992}}
  >{\raggedright\arraybackslash}p{(\columnwidth - 4\tabcolsep) * \real{0.3590}}@{}}
\toprule
\textbf{Challenge} & \textbf{What Successors Fix} & \textbf{What Remains
Unaddressed} \\
\midrule
Data Integrity & ACID transactions, schema evolution, time-travel
(Lakehouse) & Organizational discipline to use these features; data
quality at source \\
Governance & Native catalogs, lineage tracking, access control APIs
(Lakehouse/Mesh) & Cultural adoption of governance; stewardship roles;
metadata discipline \\
Ownership & Domain ownership model, data-as-product concept (Mesh) & Org
maturity to implement federated governance; domain boundary clarity \\
Skills Gap & SQL accessibility reduces barrier to entry; managed
services reduce ops burden & Data literacy; analytical thinking;
cross-functional collaboration skills \\
Culture & Minimal direct impact (architecture change $\neq$ culture change) &
Data-driven decision culture; executive sponsorship; change
management \\
\bottomrule
\end{tabular}
\end{table*}

\subsection{The Data Lakehouse: Technical Progress, Organizational Stasis}

The Data Lakehouse architecture, developed by Databricks and
subsequently adopted by major cloud providers, represents the most
commercially successful successor paradigm. The global lakehouse market
was valued at \$5.2 billion in 2023 and is projected to reach \$66.4
billion by 2033 {[}52{]}. Organizations report 35-40\% total cost of
ownership reductions compared to maintaining separate lake and warehouse
infrastructures {[}53{]}. Projections indicate that 60\% of enterprises
will adopt lakehouse solutions by 2026 {[}54{]}.

The lakehouse addresses several legitimate technical shortcomings of the
original Data Lake architecture: it provides ACID transaction support
through open table formats (Delta Lake, Apache Iceberg, Apache Hudi); it
enables direct SQL querying without requiring data movement to a
separate warehouse; and it supports both batch and streaming workloads
on a unified platform. These are genuine technical advances that
simplify architecture and reduce data duplication.

However, the lakehouse paradigm does not, and cannot, address the
organizational failures that drove Data Lake failures. Metadata drift
can still turn lakehouses into "data swamps," prompting continued
investment in automated catalogs and lineage tracking {[}53{]}. The
fundamental challenges of data quality, governance, and organizational
adoption remain. As one industry analysis notes: "54\% of executives
have made Data Governance a top priority for 2024 to 2025, but many
implementations treat governance as an afterthought" {[}55{]}. The
technology has improved; the organizational patterns have not.

\emph{C. Data Mesh: Organizational Fix Requiring Organizational
Maturity}

Data Mesh, introduced by Zhamak Dehghani in 2019 {[}56{]}, represents a
more radical departure from centralized data architecture. It explicitly
addresses organizational failures by treating data as a product owned by
domain teams, implementing federated computational governance, and
building self-serve data infrastructure. The Data Mesh market is
projected to grow from \$1.28 billion in 2023 at a CAGR of 16.3\%
through 2031 {[}57{]}. However, current adoption remains modest: surveys
show lakehouse, data fabric, and data mesh collectively have only 8-12\%
usage each {[}58{]}.

The irony of Data Mesh is that it requires precisely the organizational
capabilities that Data Lake initiatives most often lacked. According to
secondary analysis of Gartner research (as reported in {[}59{]}),
\emph{only 18\% of organizations have reached the governance maturity
level necessary to adopt the Data Mesh approach successfully}. The
approach demands domain boundary identification, technical complexity
management, skills training across distributed teams, cultural change
from centralized to decentralized control, and consistent data quality
management across autonomous domains.

Organizations that failed to implement governance for a centralized lake
are unlikely to successfully implement federated governance across
autonomous domains. As industry analysts observe: "In 2024, many
companies might move away from the concept of Data Mesh as we know it...
mainly attributed to the significant challenges organizations face when
implementing and maintaining the core principles of Data Mesh in complex
and ever-evolving business environments" {[}60{]}. Data Mesh is not
wrong in its diagnosis, but it prescribes a treatment that most patients
cannot tolerate.

\subsection{The Persistence of Failure Patterns}

\begin{figure}[!t]
\centering
\includegraphics[width=\columnwidth]{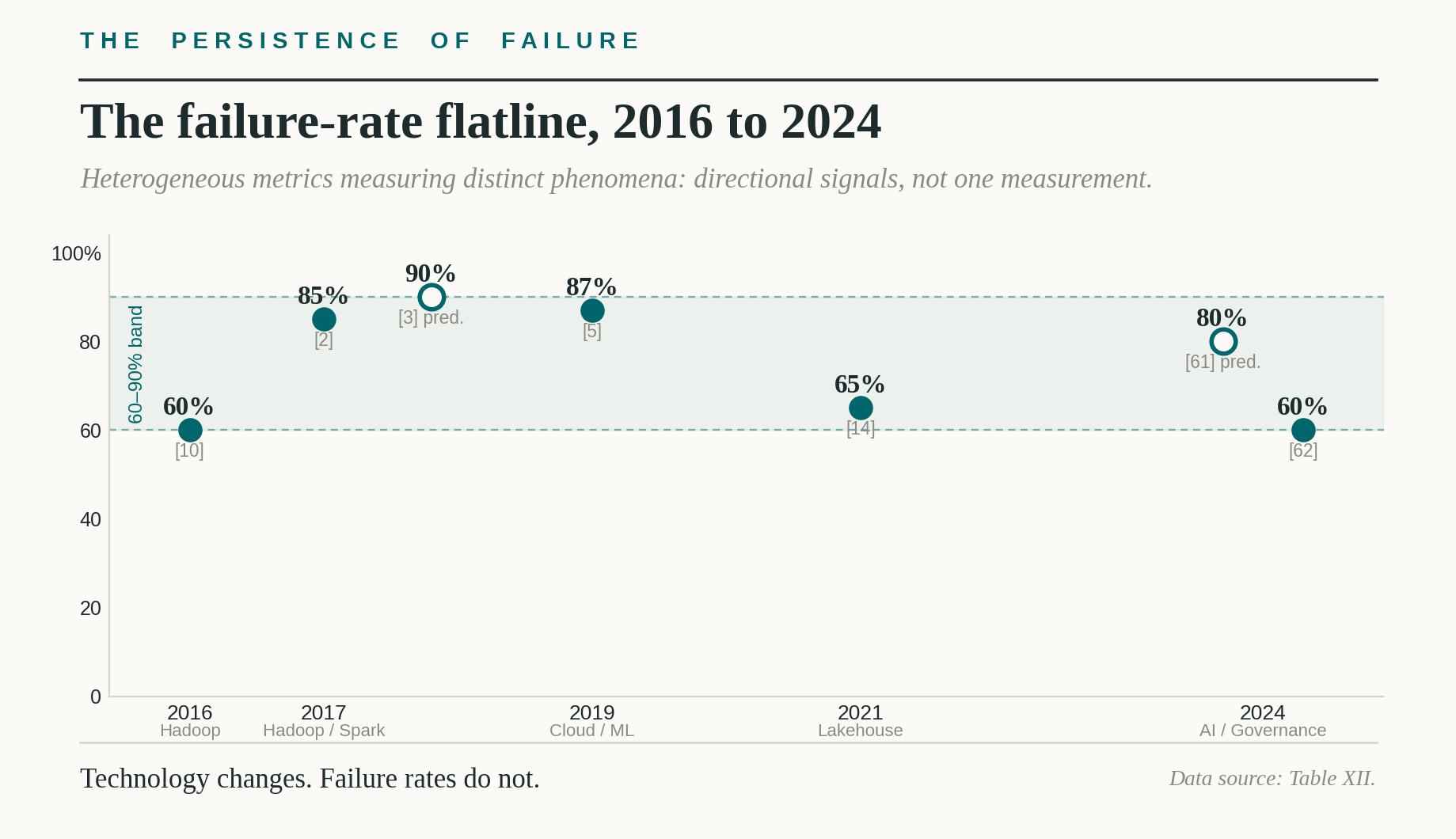}
\caption{The failure rate flatline (2016 to 2024). Heterogeneous metrics measuring distinct phenomena: directional signals, not a unified measurement. Data source: Table~\ref{tab:flat}.}
\label{fig:flat}
\end{figure}

Perhaps the most telling evidence that the industry has not learned
comes from Gartner\textquotesingle s 2024 prediction: \emph{by 2027,
80\% of data and analytics governance initiatives will fail} {[}61{]}.
This prediction, made fifteen years after Data Lakes were introduced and
after two generations of successor paradigms, suggests that the
fundamental organizational challenges remain unaddressed. The failure
rate for governance initiatives mirrors the 85\% failure rate for big
data projects reported in 2017 {[}2{]}, the numbers have not improved
(Table~\ref{tab:flat}, Fig.~\ref{fig:flat}).

Additional evidence of persistent failure patterns emerges from broader
transformation statistics. Digital transformation projects achieve only
a 35\% success rate globally, according to BCG\textquotesingle s study
of over 850 companies {[}14{]}. Organizations continue to report that
62\% cite data governance as the biggest barrier to AI adoption
{[}62{]}. Poor data costs companies 12\% of revenue, while between 60\%
and 73\% of data remains unused for any strategic purpose {[}63{]}.
Despite generational changes in technology platforms, these metrics have
remained stubbornly consistent.

\subsection{What Genuine Learning Would Look Like}

If the industry had genuinely learned from Data Lake failures, we would
expect to see several changes in how successor paradigms are adopted:

\textbf{Governance-first implementation:} Organizations would establish
governance frameworks before, not after, platform deployment. Yet
surveys consistently show governance treated as an afterthought.
Adoption of data mesh and data fabric architectures grew from 13\% in
2023 to 18\% in 2024 {[}63{]}, but this adoption frequently occurs
without the prerequisite governance maturity.

\textbf{Realistic expectation setting:} Vendor marketing would
acknowledge organizational prerequisites rather than promising
technology-driven transformation. Instead, lakehouse and mesh vendors
continue to emphasize technical capabilities while downplaying the
organizational change required.

\textbf{Investment in skills before platforms:} Organizations would
build data literacy and governance capabilities before investing in new
architectures. Yet only 28\% of organizations have achieved data
literacy despite 83\% of leaders calling it critical {[}16{]}. The
skills mirage persists.

\textbf{Cultural transformation preceding technical transformation:}
Organizations would address the cultural barriers that NewVantage
Partners has consistently identified as the primary obstacle. As
Davenport and Bean observed, firms readily embrace analytics tools yet
mostly fail to build the data-driven culture that would make those tools
pay off {[}51{]}. Instead, technology investments continue to precede
and substitute for cultural change.

\subsection{The AI Amplification Risk}

The current rush to implement generative AI and machine learning at
scale introduces new risks that amplify existing governance failures.
Gartner projects worldwide IT spending to reach \$5.6 trillion in 2025,
with AI infrastructure driving substantial growth in data center systems
{[}64{]}. Industry estimates, though derived from vendor analysis rather
than systematic research, suggest that by 2027, approximately 60\% of AI
projects may fail to realize anticipated value due to data governance
gaps {[}62{]}. The acceleration of AI adoption increases the
consequences of poor data governance: models trained on ungoverned data
inherit and amplify quality issues, producing outputs that are
unreliable, biased, or non-compliant.

The pattern repeats: organizations that have not addressed governance
debt from their Data Lake implementations are now building AI systems on
the same ungoverned data foundations. The consequences of governance
failures, already severe for analytics, become catastrophic for AI
systems that make or influence consequential decisions. The lesson of
Data Lakes, that technology cannot substitute for organizational
discipline, remains unlearned even as the stakes increase. Table XI
summarizes how each successor paradigm addresses (or fails to address)
the governance challenges identified in this analysis.

\begin{table*}[!t]
\centering
\footnotesize
\renewcommand{\arraystretch}{1.25}
\caption{COMPARATIVE ANALYSIS OF DATA ARCHITECTURE PARADIGMS}
\begin{tabular}{@{}@{}
  >{\raggedright\arraybackslash}p{(\columnwidth - 6\tabcolsep) * \real{0.2192}}
  >{\raggedright\arraybackslash}p{(\columnwidth - 6\tabcolsep) * \real{0.2536}}
  >{\raggedright\arraybackslash}p{(\columnwidth - 6\tabcolsep) * \real{0.2585}}
  >{\raggedright\arraybackslash}p{(\columnwidth - 6\tabcolsep) * \real{0.2687}}@{}}
\toprule
\textbf{Dimension} & \textbf{Data Lake} & \textbf{Data Lakehouse} &
\textbf{Data Mesh} \\
\midrule
Governance Approach & Afterthought; schema-on-read defers decisions &
Technical layer (ACID, schema enforcement); still afterthought &
Federated; requires mature governance culture \\
Maturity Required & Low (apparent); High (actual) & Medium; technical
skills sufficient & Very High; only 18\% have requisite governance
maturity {[}59{]} \\
Primary Fix & N/A (original paradigm) & Technical (ACID, SQL, unified
batch/stream) & Organizational (domain ownership, federated) \\
Adoption (2024) & Declining; legacy status & Growing rapidly; projected
mainstream by 2026 {[}54{]} & 8-12\% {[}58{]}; selective due to
complexity \\
Org. Problem Solved? & No & No (technical focus only) & Requires it as
prerequisite \\
\bottomrule
\end{tabular}
\end{table*}

\begin{table*}[!t]
\centering
\footnotesize
\renewcommand{\arraystretch}{1.25}
\caption{PERSISTENCE OF FAILURE RATES ACROSS TECHNOLOGY GENERATIONS (2015-2027)}
\label{tab:flat}
\begin{tabular}{@{}@{}
  >{\raggedright\arraybackslash}p{(\columnwidth - 6\tabcolsep) * \real{0.1466}}
  >{\raggedright\arraybackslash}p{(\columnwidth - 6\tabcolsep) * \real{0.3260}}
  >{\raggedright\arraybackslash}p{(\columnwidth - 6\tabcolsep) * \real{0.3673}}
  >{\raggedright\arraybackslash}p{(\columnwidth - 6\tabcolsep) * \real{0.1601}}@{}}
\toprule
\textbf{Year} & \textbf{Source} & \textbf{Finding} & \textbf{Rate} \\
\midrule
2016 & Gartner {[}10{]} & Big data projects will fail & 60\% \\
2017 & Gartner {[}2{]} & Big data projects will not be operationalized &
85\% \\
2017 & Gartner {[}3{]} & Data Lakes will become useless data swamps &
90\% \\
2019 & VentureBeat {[}5{]} & AI projects fail to move to production &
87\% \\
2021 & BCG {[}14{]} & Digital transformation initiatives fail & 65\% \\
2024 & Gartner {[}61{]} & D\&A governance initiatives will fail &
80\% \\
2024 & Actian {[}62{]} & AI projects fail (governance gaps) & 60\% \\
\bottomrule
\end{tabular}
\end{table*}

\textbf{Important Interpretive Note:} Table XII and Figure 4 present
failure rates from heterogeneous sources that measure \emph{distinct
phenomena} (big data project operationalization, Data Lake deployment
utility, AI/ML productionization, digital transformation success,
governance initiative effectiveness) using \emph{different definitions
of \textquotesingle failure\textquotesingle{}} (project abandonment,
failure to reach production, failure to deliver ROI, failure to achieve
organizational change). These rates derive from varied methodologies
(analyst estimates, self-reported surveys, market research) with
different sample populations. They are \emph{not directly comparable and
should not be averaged or aggregated}. We present them not as a unified
metric but as convergent signals: when multiple independent sources,
measuring different aspects of data initiative performance, consistently
report high failure rates (60-90\%) over a decade, the pattern warrants
investigation regardless of measurement heterogeneity. The persistence
of high rates across technological generations, not the specific
percentages, is the analytically significant finding.

\section{Field Evidence: A Practitioner's Reality-Check Catalogue}
The preceding sections reconstruct what the industry said. This section reports what I observed directly. The evidence I bring is not a set of metrics: it is a catalogue of close to five hundred field ``reality checks'' that I recorded over fifteen years of designing, delivering, and rescuing enterprise Data Lake platforms, predominantly in banking, consumer finance, and telecommunications. Each entry is a short statement of something that goes wrong, written the way I encountered it on the ground. The full catalogue is provided as supplementary material; this section reports what the corpus reveals when I step back from it.

\subsection{Why a Catalogue, and How to Read It}
I deliberately offer qualitative practitioner knowledge rather than quantitative claims. Failure counts drawn from confidential client engagements would be neither verifiable nor, in my view, honest: they would dress recollection as measurement. A curated catalogue is more defensible. It is the accumulated, dated residue of real programs, and it can be read as the negative image of governance: every entry names a discipline that was absent. I coded the catalogue thematically against the framework developed earlier in this paper. The exercise is itself a finding. The catalogue was assembled from practice, independently of the analyst literature, yet it maps almost entirely onto the Seven Deadly Sins and the components of Governance Debt. That convergence, practitioner corpus meeting independent literature, is the strongest validation I can offer, and it requires no statistics. I also acknowledge my reflexive position: I was an actor in many of these situations, and several of these reality checks I learned by committing the error myself before learning to name it.

For this paper I consolidated the close to five hundred dated entries by collapsing duplicates and near-duplicates, then sorted the remainder into sixteen thematic clusters, each tied to one of the Seven Deadly Sins, one component of Governance Debt, or Governance Gravity. Tables~\ref{tab:cat1} and~\ref{tab:cat2} make the corpus auditable: for every cluster they give representative reality checks, phrased as I recorded them on the ground, alongside the framework element the cluster maps to. I report no per-cluster counts. That is deliberate, and consistent with the decision above to treat the catalogue as a qualitative corpus rather than a source of metrics.

\begin{table*}[!t]
\centering
\footnotesize
\renewcommand{\arraystretch}{1.25}
\caption{THE PRACTITIONER CATALOGUE MAPPED TO THE FRAMEWORK (CLUSTERS 1 TO 8)}
\label{tab:cat1}
\begin{tabular}{@{}>{\raggedright\arraybackslash}p{0.20\textwidth} >{\raggedright\arraybackslash}p{0.52\textwidth} >{\raggedright\arraybackslash}p{0.21\textwidth}@{}}
\toprule
\textbf{Cluster} & \textbf{Representative field reality checks (recorded as encountered)} & \textbf{Maps to} \\
\midrule
Purpose, scoping, and ingestion & Launching the lake before any inventory of source systems or data dictionary; ``ingest everything, understand later,'' with no use case and no prioritization. & Sin 1, Gluttony \\
Structure, zoning, and conventions & Raw, cleansed, curated, and serving data left in one place; no naming, partitioning, or versioning conventions; schema drift unmanaged. & Sin 2, Sloth \\
Metadata, cataloguing, and lineage & No documentation of datasets, owners, or freshness; a catalogue filled once by hand then abandoned, with no lineage to trace where a datum came from. & Governance Debt: metadata \\
Data quality, reconciliation, and trust & Quality deferred to the end of the pipeline or never measured; unexplained divergence between warehouse, lake, and Excel, with no golden records or matching rules. & Governance Debt: quality; feeds Governance Gravity \\
Governance and operating model & No owners, stewards, or RACI at launch; governance on paper only, committees that decide nothing, owners named without authority or dedicated time. & Sin 3, Pride; Governance Debt: organizational \\
Security, access, and compliance & No classification of personal or regulated data; broad access by convenience and shared accounts; security and privacy bolted on after the fact rather than designed in. & Governance Debt: security and compliance \\
Lifecycle, retention, and obsolescence & No retention or purge rules, everything kept ``just in case''; no distinction between hot, warm, and cold data; no knowledge of which datasets are dead. & Governance Debt: quality and cost \\
Pipeline robustness and integration & No SLAs, load windows, or dependency management between flows; no handling of rejects, retries, idempotence, late-arriving data, or backfills. & Reliability; engineering-discipline debt \\
\bottomrule
\end{tabular}
\end{table*}

\begin{table*}[!t]
\centering
\footnotesize
\renewcommand{\arraystretch}{1.25}
\caption{THE PRACTITIONER CATALOGUE MAPPED TO THE FRAMEWORK (CLUSTERS 9 TO 16)}
\label{tab:cat2}
\begin{tabular}{@{}>{\raggedright\arraybackslash}p{0.20\textwidth} >{\raggedright\arraybackslash}p{0.52\textwidth} >{\raggedright\arraybackslash}p{0.21\textwidth}@{}}
\toprule
\textbf{Cluster} & \textbf{Representative field reality checks (recorded as encountered)} & \textbf{Maps to} \\
\midrule
Engineering discipline and DataOps & No versioning of code, configurations, or schemas and no data CI/CD; hardcoded credentials and paths; black-box pipelines and scripts nobody can safely change. & Engineering-discipline debt (proposed) \\
Operations, observability, and the run & No supervision, alerting, or runbooks and no funded run team; sizing on storage volume alone; no workload isolation, so a single job starves the cluster. & Operational debt (proposed) \\
Technology choices and tool stacking & Choosing technology before architecture, operating model, or skills; stacking Atlas, Ranger, NiFi, Spark, and Kafka with no model of use; streaming with no real-time need. & Sin 4, Idolatry \\
Adoption, value, and change management & Self-service BI over ungoverned indicators; no enablement, portal, or semantic layer; Excel remaining the real platform; no measure of adoption or reuse. & Sin 5, Envy \\
Skills, knowledge, and dependency & Total dependence on the integrator with no knowledge transfer; no documentation of delivered pipelines; critical business rules known to a single person. & Sin 6, Greed \\
Cost and the economics of the run & Reasoning on infrastructure cost only while ignoring total cost of ownership; open source treated as free; the run never funded, so true costs surface after go-live. & Sin 7, Wrath \\
Coexistence with the warehouse & No migration or coexistence plan; the lake presented as a replacement for everything, then the discovery that the warehouse cannot be decommissioned. & Governance Gravity \\
Underestimated legacy and hidden complexity & Business rules buried in scripts, hidden processing on desktops, files sent by email; underestimated entity matching, historization, and partner-versus-internal data trust. & The Legacy Amplifier \\
\bottomrule
\end{tabular}
\end{table*}

\subsection{The Catalogue Mapped to the Seven Sins}
The full cluster-by-cluster mapping appears in Tables~\ref{tab:cat1} and~\ref{tab:cat2}. Read against the taxonomy, the corpus is unambiguous. \emph{Gluttony} dominates the opening of almost every troubled program: launching a lake without a serious source inventory, without a data dictionary, ``ingest everything, understand later,'' confusing having the data with understanding it. \emph{Sloth} follows in the refusal to structure: failing to separate Raw, Cleansed, Curated, and Serving zones, no naming or partitioning conventions, schema drift left unmanaged. \emph{Pride} appears as governance bolted on after the fact: PowerPoint governance, committees that decide nothing, data owners named without authority or dedicated time. \emph{Idolatry} runs through the catalogue as tool stacking and fashion-driven choice: installing Atlas, Ranger, Airflow, NiFi, Spark, and Kafka with no model of use, moving from Hadoop to cloud to Lakehouse without resolving the real problems, ``Big Data without big data.'' \emph{Envy} surfaces as the democratization illusion: self-service BI over ungoverned indicators, opening everything without control. \emph{Greed} is the skills mirage and the dependency it breeds: total reliance on the integrator, black-box pipelines, scripts nobody understands, the silent risk of the one expert who leaves. \emph{Wrath} closes the cycle as the cost delusion: believing cheap storage justifies massive storage, never funding the run, assuming the project ends at go-live.

\subsection{What the Catalogue Adds to the Literature}

\begin{figure*}[!t]
\centering
\includegraphics[width=\textwidth]{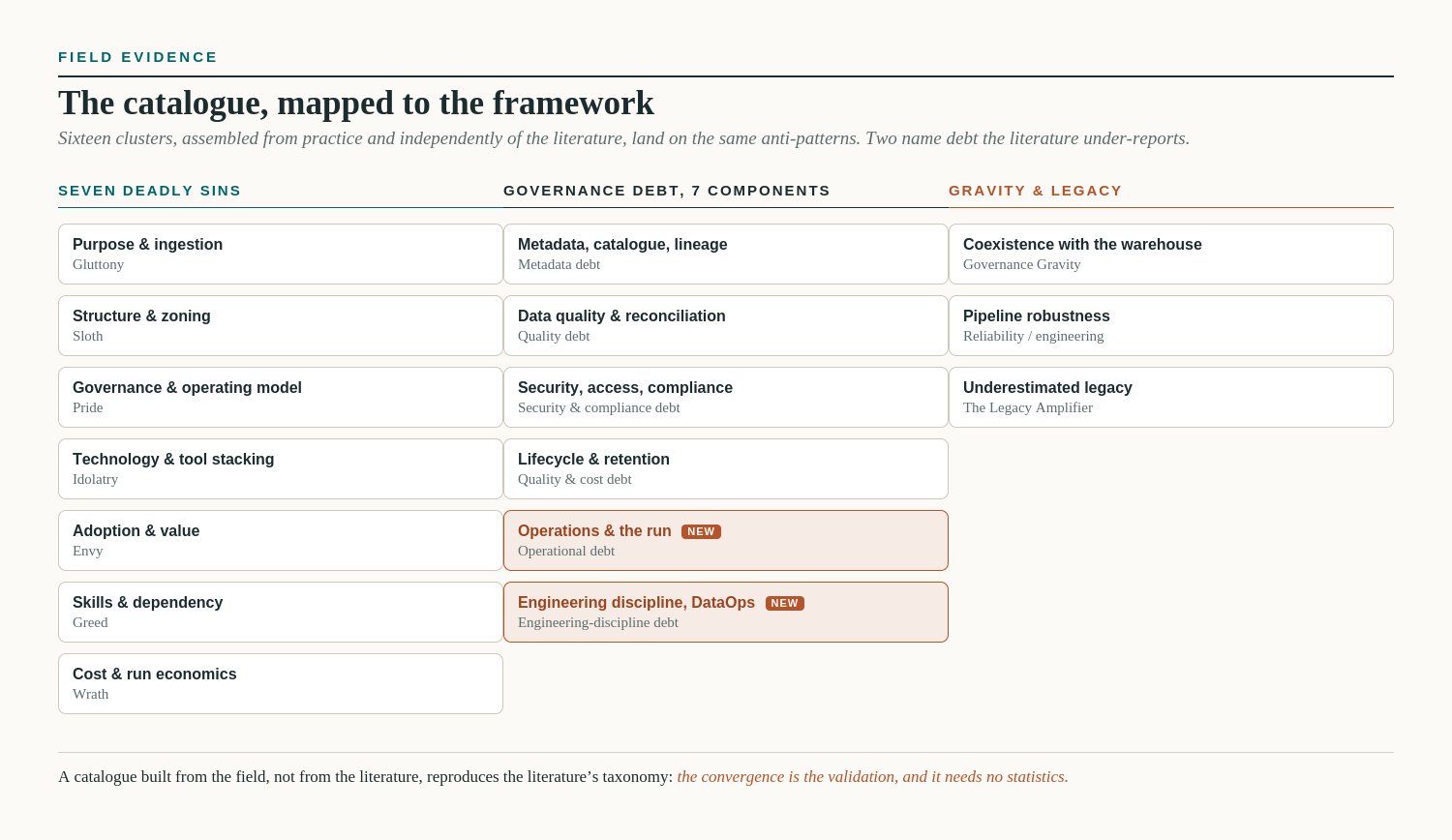}
\caption{The field catalogue mapped to the framework. Sixteen clusters, assembled from practice and independently of the literature, land on the Seven Deadly Sins, the components of Governance Debt, and Governance Gravity. Two clusters name debt the literature under-reports: operational debt and engineering-discipline debt.}
\label{fig:conv}
\end{figure*}

Two clusters in my catalogue are heavily populated yet under-weighted in the analyst and academic record, and I treat them as a contribution rather than a footnote (Fig.~\ref{fig:conv}). The first is \emph{operational debt}: the failures of the run. Pipelines that are not supervised, no observability, no alerting, no runbooks, workloads that are not isolated so a single Spark job starves the cluster, capacity planned without knowing the workloads, and the recurring belief that delivery ends at production. The literature treats governance largely as cataloguing, lineage, and access; my field record insists that a lake dies just as often from operational neglect as from missing metadata. The second is \emph{engineering-discipline debt}: no versioning of code, configurations, schemas, or datasets, no data CI/CD, no tests on transformations, no before-and-after reconciliation at migration, credentials and paths hardcoded. These are not glamorous failures, which is precisely why they are absent from vendor narratives and present in every post-mortem I have written. I therefore propose adding operational debt and engineering-discipline debt as first-class components of Governance Debt alongside metadata, quality, security, compliance, and organizational debt.

\subsection{The Catalogue Confirms Governance Gravity}
Finally, a thick seam of the catalogue speaks directly to Governance Gravity, and from the field rather than from theory. The reality checks recur: divergences between the warehouse, the lake, and Excel that no one can explain; reconciliation treated as optional; golden records and master data never defined; the lake quietly positioned as a replacement for everything; Excel remaining the real platform because the metiers never stopped extracting, reworking, and correcting locally. I did not derive Governance Gravity from a model. I watched organizations return to the warehouse the moment a regulatory or executive figure could not be reproduced, and the catalogue is the trace those retreats left behind.

\section{The Moroccan and African Context}
Most of the failure literature is written from a North American or European enterprise vantage. My own work sits in an emerging-market context whose conditions do not merely replicate that experience: they amplify several of the anti-patterns. I identify four contextual forces and then map them, in Table~\ref{tab:context}, onto the specific sins they intensify.

\emph{Talent scarcity and concentration.} Specialized data engineering and governance skills are scarcer and far more concentrated than in mature markets. A handful of senior profiles anchor most serious programs, and competition from offshore and overseas employers accelerates their turnover. The consequence is twofold. The Skills Mirage is more acute, because the institutional knowledge needed to interpret raw lake data is thinner at the outset. And the dependency it breeds is heavier, because when that knowledge leaves it is rarely replaced at equivalent depth, so metadata and engineering-discipline debt accumulate faster than in markets with a deeper labor pool.

\emph{Regulatory regime.} Data handling is governed by Morocco's Law 09-08 on the protection of personal data and supervised by the CNDP, while banking platforms operate under Bank Al-Maghrib oversight. The regime is real and tightening, but its enforcement history and its localization expectations differ from those of GDPR. In practice this shapes the compliance and security components of governance debt in a distinctive way: controls are often retrofitted to satisfy a supervisory request rather than designed in, and data localization constraints narrow the set of viable cloud architectures from the outset.

\emph{Vendor and infrastructure ecosystem.} Cloud region availability, connectivity, procurement cycles, and the local partner ecosystem differ markedly from mature markets. Managed services that elsewhere absorb operational burden are not always available locally or contractually, which pushes organizations toward self-managed stacks and inflates the very run costs the Cost Delusion ignores. Procurement and licensing cycles also lengthen the time between a decision and a working capability, encouraging the platform-first reflex.

\emph{Organizational and linguistic factors.} Programs depend heavily on systems integrators, technical documentation and tooling are often consumed in French while products are documented in English, and the relationship to data is frequently more hierarchical than self-service. These factors interact with the cultural and organizational components of governance debt and make the Democratization Illusion particularly tempting, since self-service is announced long before the literacy and governance that would make it real.

The contribution here is not a claim of exceptionalism but a documented account of how universal anti-patterns express themselves under emerging-market constraints, a perspective largely absent from the existing record. Table~\ref{tab:context} makes the amplification explicit.

\begin{table*}[!t]
\centering
\footnotesize
\renewcommand{\arraystretch}{1.3}
\caption{How Emerging-Market Conditions Amplify Specific Anti-Patterns}
\label{tab:context}
\begin{tabular}{@{}p{3.4cm}p{6.2cm}p{6.4cm}@{}}
\toprule
\textbf{Contextual force} & \textbf{Local condition} & \textbf{Anti-pattern and debt it amplifies} \\
\midrule
Talent scarcity and concentration & Few senior profiles; fast turnover toward offshore employers & Skills Mirage (Sin 6); accelerates metadata and engineering-discipline debt as knowledge departs \\
Regulatory regime & Law 09-08 / CNDP and Bank Al-Maghrib oversight, with distinct enforcement maturity and localization expectations & Governance as Afterthought (Sin 3); shapes compliance and security debt, controls retrofitted to requests \\
Vendor and infrastructure ecosystem & Limited managed-service and cloud-region availability; long procurement cycles & Cost Delusion (Sin 7) and Technology Worship (Sin 4); inflates run cost, encourages platform-first reflex \\
Organizational and linguistic factors & Integrator dependence, francophone working language, hierarchical data culture & Democratization Illusion (Sin 5); deepens organizational debt and dependency \\
\bottomrule
\end{tabular}
\end{table*}

\section{DISCUSSION AND REALITY CHECK FRAMEWORK}

One finding runs through this retrospective and cuts against most of the
discourse, academic and practitioner alike: \emph{Data Lake failures
were organizational long before they were technological.} Keep treating
an organizational problem as a technical one and the failure rate stays
where it has been for fifteen years, regardless of which platform is in
fashion.

\subsection{Synthesis of Findings}

Our analysis addressed the four uncomfortable questions posed in the
introduction. First, regarding why failure rates remained persistently
high: the evidence reviewed across 64 sources suggests that
organizations systematically underestimated organizational prerequisites
while overestimating technological capabilities. The "Seven Deadly Sins"
we identified, purposeless ingestion, schema avoidance, governance as
afterthought, technology worship, democratization illusion, skills
mirage, and cost delusion, appear to represent organizational
pathologies rather than technical limitations, based on the patterns
observed across practitioner and analyst accounts.

Second, concerning whether original promises were realistic: the
reviewed evidence suggests they were not. Each promise embedded implicit
assumptions about organizational capabilities that most enterprises did
not possess. The assumption that schema-on-read would enable flexibility
assumed organizations could impose schema discipline at consumption
time, as reported across multiple sources, they typically could not. The
assumption that democratization would follow from access assumed data
literacy that did not exist. The assumption that commodity
infrastructure would reduce costs ignored the total cost of ownership
including scarce human capital.

Third, regarding governance debt: we propose that deferred governance
decisions accumulate compound interest through knowledge degradation,
propagation effects, regulatory ratchet, and volume amplification. The
governance debt framework provides a conceptual tool for understanding
why remediation becomes progressively more expensive and eventually
impossible, a hypothesis that awaits formal empirical validation.

\begin{figure*}[!t]
\centering
\includegraphics[width=\textwidth]{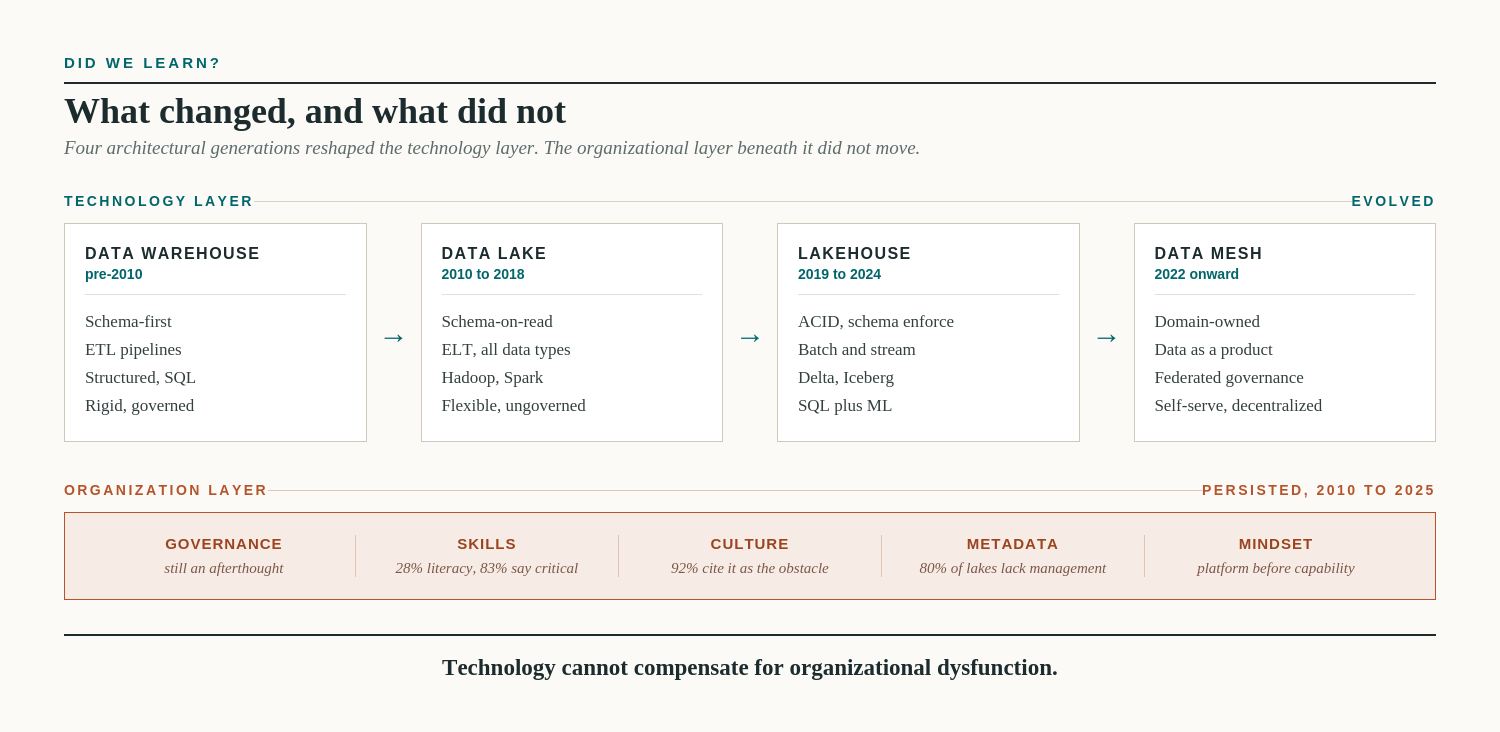}
\caption{What changed and what did not. Four architectural generations reshaped the technology layer; the organizational layer beneath it persisted unchanged.}
\label{fig:evo}
\end{figure*}

Fourth, concerning whether successor paradigms learned from these
failures: the available evidence suggests limited learning.
Gartner\textquotesingle s 2024 prediction that 80\% of governance
initiatives will fail by 2027 {[}61{]} mirrors the 85\% failure rate for
big data projects reported in 2017 {[}2{]}. As reflected in our
comparative analysis, the fundamental organizational challenges appear
to persist across technological generations (Fig.~\ref{fig:evo}).

\subsection{Boundary Conditions: When Data Lakes Succeed}

The critical perspective of this analysis should not obscure that Data
Lakes have delivered value for some organizations. Understanding the
boundary conditions that distinguish success from failure strengthens
rather than undermines our central thesis: organizational factors, not
technology, determine outcomes. Organizations that succeed with Data
Lakes typically exhibit several characteristics that directly address
the anti-patterns we identified.

\textbf{Success Pattern 1: Governance-First Culture.} Successful
implementations treat governance as a prerequisite, not an afterthought.
These organizations establish data ownership, quality standards, and
metadata requirements before scaling ingestion. They accept what
practitioner accounts suggest is approximately 10-15\% overhead for
`governance gates' at Stage 1 (Table~\ref{tab:stage}), a heuristic estimate requiring
organizational calibration, because they recognize the exponential cost
of remediation at later stages. Data-native companies (born-digital
organizations with engineering-led cultures) often exhibit this pattern
naturally.

\textbf{Success Pattern 2: Use-Case-Driven Ingestion.} Rather than
\textquotesingle ingest everything, analyze later,\textquotesingle{}
successful implementations require articulated use cases before data
flows into the lake. Each dataset has a named consumer, defined purpose,
and documented quality requirements. This directly counters
\textquotesingle Sin 1: Purposeless Ingestion\textquotesingle{} and
prevents the accumulation of undocumented, unused datasets that
characterize swamp formation.

\textbf{Success Pattern 3: Skills Investment Parallel to Platform
Investment.} Organizations that succeed invest in data literacy,
engineering capability, and governance expertise alongside
infrastructure. They recognize that democratization requires enablement,
not just access. Budget allocations reflect this: based on patterns
observed in practitioner accounts, successful programs dedicate an
estimated 30-40\% of investment to capability building rather than the
10-15\% range commonly reported in failed initiatives (these proportions
are indicative heuristics, not empirically validated benchmarks).

\textbf{Success Pattern 4: Explicit Coexistence Strategy.} Rather than
pursuing aggressive warehouse replacement, successful organizations
acknowledge governance gravity and plan for long-term coexistence. They
position the lake for use cases where its strengths (schema flexibility,
cost-effective storage, diverse data types) provide genuine advantage,
while maintaining the warehouse as trusted substrate for regulatory,
financial, and executive workloads. This pragmatic positioning avoids
the trust erosion that triggers reversion.

These success patterns reinforce our central argument: technology
succeeds when organizational foundations are in place. The lake itself
is not flawed, the premise that technology alone could overcome
organizational deficits was flawed. Organizations that treat the lake as
one component of a broader data management strategy, with appropriate
governance, skills, and realistic expectations, can and do extract
substantial value.

\subsection{The Technology-Organization Gap}

A persistent gap exists between technological capability and
organizational capacity to exploit that capability. This gap manifests
in several ways. Technologies are adopted faster than organizations can
develop the skills, processes, and culture to use them effectively.
Vendor promises emphasize technical features while downplaying
organizational prerequisites. Academic research focuses
disproportionately on technical architectures rather than sociotechnical
implementation challenges. Investment decisions prioritize platform
acquisition over capability building.

This gap is not unique to Data Lakes, it characterizes technology
adoption broadly, but the Data Lake case is instructive because it
involves the foundational asset of data. When organizations fail to
govern their data infrastructure, the consequences cascade through every
subsequent initiative: analytics, machine learning, AI, and digital
transformation all inherit the dysfunction of the underlying data
foundation.

\subsection{Implications for Research}

Our analysis suggests several directions for future research. First, the
academic literature on data management architectures would benefit from
greater attention to organizational factors. Technical surveys of Data
Lake architectures are abundant; empirical studies of organizational
adoption challenges are scarce. Research methodologies from information
systems, organizational behavior, and technology adoption should be
applied to understand why technically sound architectures fail in
practice.

Second, the governance debt concept introduced in this paper warrants
further development. Quantitative methods for measuring governance debt,
empirical validation of compound interest effects, and case studies of
remediation efforts would contribute to both theory and practice. The
relationship between technical debt and governance debt, how they
interact and compound, remains unexplored.

Third, longitudinal studies tracking the same organizations through
multiple technology generations would illuminate whether and how
organizational learning occurs. The pattern of repeated failure across
Hadoop, cloud data lakes, and lakehouses suggests that organizational
dysfunction persists independently of technology choices, but this
hypothesis requires systematic investigation.

\subsection{Implications for Practice}

For practitioners, our analysis counsels humility and realism. The
reported failure rates, ranging from 60\% to 90\% across different
studies measuring different phenomena, have remained stubbornly
consistent across fifteen years of technological evolution, despite
their methodological heterogeneity. This convergent pattern should
dispel any expectation that new platforms will succeed where previous
ones failed, absent fundamental organizational change. Several practical
implications follow.

Organizations should assess their governance maturity before, not after,
selecting data platforms. The finding that only 18\% of organizations
possess the governance maturity required for Data Mesh adoption applies
broadly: most organizations lack the prerequisites for sophisticated
data architectures regardless of the specific paradigm chosen.

Investment in skills and culture should precede investment in platforms.
The persistent finding that 92\% of executives cite culture as the
primary barrier to becoming data-driven indicates where attention should
focus. Technology procurement without corresponding investment in human
capability is a recipe for expensive failure.

Governance should be treated as foundational infrastructure, not
optional overhead. The governance debt framework demonstrates that
deferred governance becomes progressively more expensive. Organizations
should budget for governance as they budget for security: as
non-discretionary investment in risk management.

\subsection{Reality Check Framework for Practitioners}

Based on the patterns identified in this analysis, we propose a Reality
Check Framework (Table~\ref{tab:rcf}) to help organizations assess their
readiness for data platform initiatives. The framework synthesizes the
Seven Deadly Sins and Governance Debt Assessment Model into practical
evaluation dimensions. Organizations should conduct this assessment
before platform selection, not after deployment.

\begin{table*}[!t]
\centering
\footnotesize
\renewcommand{\arraystretch}{1.25}
\caption{REALITY CHECK FRAMEWORK: PRE-IMPLEMENTATION ASSESSMENT}
\label{tab:rcf}
\begin{tabular}{@{}@{}
  >{\raggedright\arraybackslash}p{(\columnwidth - 4\tabcolsep) * \real{0.2749}}
  >{\raggedright\arraybackslash}p{(\columnwidth - 4\tabcolsep) * \real{0.3824}}
  >{\raggedright\arraybackslash}p{(\columnwidth - 4\tabcolsep) * \real{0.3427}}@{}}
\toprule
\textbf{Dimension} & \textbf{Key Questions} & \textbf{Red Flags (Stop
Signs)} \\
\midrule
Purpose Clarity & What specific business outcomes justify this
initiative? Who are the defined consumers? & "Store now, analyze later";
no defined use cases; technology-driven rationale \\
Governance Readiness & Is governance framework defined before platform
selection? Are data stewards assigned? & "Governance as Phase 2"; no
ownership model; compliance as afterthought \\
Skills \& Literacy & Do target users have requisite data literacy? Is
training budgeted and scheduled? & Assuming "access = usage"; no skills
assessment; training deferred \\
Cultural Alignment & Does leadership model data-driven decision making?
Is there executive sponsorship? & IT-only initiative; no business
sponsor; culture cited as "soft" concern \\
Total Cost Realism & Does TCO include integration, skills, governance,
and ongoing operations? & License-only budgeting; "we\textquotesingle ll
figure it out"; hidden cost assumptions \\
Success Metrics & Are business outcome KPIs defined? How will value
realization be measured? & Technical metrics only (data volume,
queries); no business outcomes defined \\
\bottomrule
\end{tabular}
\end{table*}

Organizations exhibiting multiple red flags should address these
organizational prerequisites before proceeding with platform investment.
Similarly, organizations with existing Data Lakes should assess swamp
indicators (Section V): the presence of three or more indicators,
absent ownership, incomplete metadata, unused datasets, undocumented
lineage, excessive duplication, or missing quality validation, signals
that remediation must precede any new initiative. Continuing to build on
a swamp foundation guarantees the replication of failure patterns. The
framework operationalizes the core insight of this analysis: technology
succeeds only when organizational foundations are in place.

For organizations considering data warehouse decommissioning, the
governance gravity mechanisms identified in Section VI.D translate into
specific readiness criteria. Table~\ref{tab:decom} operationalizes these mechanisms
into assessable prerequisites:

\begin{table*}[!t]
\centering
\footnotesize
\renewcommand{\arraystretch}{1.25}
\caption{DATA WAREHOUSE DECOMMISSION READINESS CHECKLIST}
\label{tab:decom}
\begin{tabular}{@{}@{}
  >{\raggedright\arraybackslash}p{(\columnwidth - 4\tabcolsep) * \real{0.2862}}
  >{\raggedright\arraybackslash}p{(\columnwidth - 4\tabcolsep) * \real{0.2550}}
  >{\raggedright\arraybackslash}p{(\columnwidth - 4\tabcolsep) * \real{0.4587}}@{}}
\toprule
\textbf{Readiness Criterion} & \textbf{Mechanism Addressed} &
\textbf{Indicative Threshold} \\
\midrule
Semantic contract parity & Semantic anchoring & Certified glossary
covers $\geq$90\% critical KPIs \\
Audit trail \& lineage & Institutional trust & Automated lineage for all
regulatory reports \\
Ownership \& stewardship & Institutional trust & 100\% critical datasets
with designated owner \\
Change control process & Operating model inertia & Documented change
management active \\
SLA equivalence & Operating model inertia & Lake SLAs $\geq$ DWH SLAs for
migrated workloads \\
Reconciliation rules & Semantic anchoring & Automated reconciliation
with alerting \\
Retention \& legal hold & Institutional trust & Retention policies +
legal hold + evidence pack replicated \\
\bottomrule
\end{tabular}
\end{table*}

Organizations failing to meet these criteria should treat the data
warehouse as the continuing source of truth for affected workloads,
positioning the lake as complementary infrastructure for exploratory
analytics, data science, and ML workloads where governance requirements
differ. Attempting decommissioning without governance parity recreates
the conditions for governance gravity to reassert itself.

\textbf{Stage-Based Intervention Strategy:} The degradation pathway
analysis (Section VI.D) implies differentiated interventions depending on
current stage. Table~\ref{tab:stage} provides a playbook mapping stages to
appropriate controls, expected effort, and anticipated impact. This
framework operationalizes the key insight that prevention (Stages 1-2)
requires fundamentally different approaches than remediation (Stage 3)
or strategic repositioning (Stages 4-5).

\begin{table*}[!t]
\centering
\footnotesize
\renewcommand{\arraystretch}{1.25}
\caption{STAGE-BASED INTERVENTION MATRIX}
\label{tab:stage}
\begin{tabular}{@{}@{}
  >{\raggedright\arraybackslash}p{(\columnwidth - 8\tabcolsep) * \real{0.2088}}
  >{\raggedright\arraybackslash}p{(\columnwidth - 8\tabcolsep) * \real{0.1603}}
  >{\raggedright\arraybackslash}p{(\columnwidth - 8\tabcolsep) * \real{0.2863}}
  >{\raggedright\arraybackslash}p{(\columnwidth - 8\tabcolsep) * \real{0.1767}}
  >{\raggedright\arraybackslash}p{(\columnwidth - 8\tabcolsep) * \real{0.1678}}@{}}
\toprule
\textbf{Stage} & \textbf{Strategy} & \textbf{Key Controls} &
\textbf{Effort} & \textbf{Expected Impact} \\
\midrule
Stage 1 (Enthusiasm) & Prevention gates & Mandatory metadata on ingest;
ownership assignment required; data contracts for new sources;
DQ-as-code in CI/CD & Low (10-15\% overhead) & Prevents swamp
formation \\
Stage 2 (Accumulation) & Governance retrofit & Catalog backfill sprint;
lineage documentation; stewardship assignment; quality baseline
establishment & Medium (1-2 FTE quarters) & Arrests degradation; debt
stabilized \\
Stage 3 (Degradation) & Targeted remediation & Dataset triage
(keep/archive/delete); certification program for high-value assets;
stewardship sprint; quality incident SLAs & High (dedicated team 6-12
months) & Partial recovery; critical assets salvaged \\
Stage 4 (Reversion) & Parallel strategy & Accept DW for trusted
workloads; lake for exploration/ML; formal coexistence architecture;
decommission abandoned datasets & Medium (architectural refactoring) &
Stable coexistence; cost optimization \\
Stage 5 (Abandonment risk) & Rebuild vs repair decision & Cost-benefit
analysis; greenfield evaluation; governance-first architecture; lessons
learned documentation & Very high (strategic initiative) & Fresh start
with embedded governance \\
\bottomrule
\end{tabular}
\end{table*}

The intervention matrix makes explicit that effort required grows
non-linearly with stage progression, the estimated `10-15\% overhead' of
Stage 1 prevention gates represents orders of magnitude less investment
than the `strategic initiative' required at Stage 5. Note that all
effort estimates in Table~\ref{tab:stage} are heuristics derived from practitioner
experience patterns and should be calibrated to organizational context;
they are indicative rather than prescriptive. This asymmetry reinforces
the compound interest nature of governance debt and underscores why
early intervention delivers disproportionate returns.

\subsection{Limitations}

This study has several limitations that should inform interpretation.
First, the retrospective methodology relies substantially on industry
analyst reports, practitioner accounts, and survey data rather than
primary empirical research. While we triangulated across multiple source
types, the underlying data reflects self-reported assessments that may
be subject to various biases.

Second, survivor bias affects the available evidence. Organizations that
abandoned Data Lake initiatives without public documentation leave fewer
traces than those with published accounts. The actual failure rate may
exceed reported figures.

Third, the governance debt framework, while conceptually grounded in
technical debt literature, has not been empirically validated. The
compound interest metaphor provides explanatory value, but the specific
dynamics of governance debt accumulation require quantitative
investigation.

Fourth, source quality varies across the evidence base. While we
prioritized peer-reviewed academic publications and established analyst
firms (Gartner, McKinsey, Forrester, NewVantage Partners) for core
claims, some supporting statistics derive from vendor white papers and
market research firms whose methodologies are not fully disclosed. We
have attempted to triangulate such claims across multiple sources and
have flagged predictions as directional assessments rather than
empirical measurements. Readers should apply appropriate skepticism to
specific percentages while recognizing that directional patterns
converge across source types: high failure rates, persistent governance
challenges, and organizational primacy over technology.

Finally, this analysis focuses on organizational failure patterns
without fully accounting for successful implementations. Data Lakes have
delivered value for some organizations, particularly data-native
companies with established engineering cultures. A more complete
analysis would identify factors that distinguish successful
implementations from unsuccessful ones.

\textbf{Comprehensive Note on Numeric Thresholds:} This analysis
proposes several numeric thresholds that require explicit
qualification. Those that remain in the main text, the swamp indicators
(\textgreater50\% ownership gap, \textless40\% metadata,
\textgreater40\% dormancy, \textgreater20\% duplication), the governance
gravity signals (\textgreater70\%, \textgreater80\%), and the Stage 1
prevention overhead (10-15\%), are \emph{heuristics derived from
practitioner experience and pattern synthesis, not empirically validated
cutoffs}. The GDAM itself carries no numeric score: it is read
qualitatively across its five dimensions (Section VI.G). The more
speculative numeric instruments, the governance parity-gap metrics and
the swamp severity heuristic, are deliberately confined to Appendix A for
the same reason. These thresholds should be treated as: (a) starting points for
organizational calibration rather than universal standards; (b) testable
hypotheses for future empirical research rather than prescriptive rules;
(c) directionally indicative rather than precisely predictive; and (d)
subject to variation by industry, organization size, regulatory
intensity, and data maturity. The preliminary expert review (Section
V.G) established face validity for these heuristics, but formal
empirical validation remains necessary. Organizations adopting these
frameworks should track their own threshold-to-outcome correlations and
refine thresholds based on observed predictive accuracy in their
context. \textbf{The contribution of this analysis lies in the
measurement system and trajectory tracking methodology, not in the
specific numeric cutoffs}. Deteriorating metrics over consecutive
quarters signal governance degradation regardless of whether
organizations breach the proposed thresholds precisely. The directional
pattern matters more than specific values.

\textbf{Threats to Validity:} Several specific threats warrant explicit
acknowledgment. (1) \emph{Construct validity}: the operational
definitions of \textquotesingle failure,\textquotesingle{}
\textquotesingle swamp,\textquotesingle{} and
\textquotesingle governance debt\textquotesingle{} vary across sources;
our synthesis necessarily imposes conceptual coherence that may obscure
meaningful distinctions. (2) \emph{Internal validity}: the causal
mechanisms proposed (governance gravity, legacy amplifier) are inferred
from correlational patterns rather than controlled observation;
alternative explanations cannot be ruled out. (3) \emph{External
validity}: the evidence base is skewed toward large enterprises in
regulated industries; patterns may not generalize to SMEs, startups, or
less regulated sectors. (4) \emph{Threshold reliability}: the numeric
thresholds that remain, the swamp indicators and the governance
gravity signals (\textgreater70\%, \textgreater80\%), are heuristics
derived from synthesis rather than empirically calibrated cutoffs, and
the more speculative numeric instruments are confined to Appendix A; all
should be treated as hypotheses for future validation rather than
prescriptive standards. (5)
\emph{Source selection bias}: our evidence base relies substantially on
gray literature (industry reports, vendor white papers, practitioner
blogs) that may reflect survivorship bias, promotional interests, or
selective reporting. Gray literature sources are more likely to report
failures that generate engagement than unremarkable successes. We
mitigate this through cross-referencing across source types and treating
claims as directional signals rather than precise measurements.
\emph{Practitioner-corpus reflexivity}: the field catalogue is the
record of a single practitioner who was an actor in many of the
situations it describes. This grants first-hand access that external
reviewers lack, but it also introduces selection effects (the author
chose what was worth noting), hindsight, and a perspective weighted
toward the sectors and geography of his engagements. The catalogue is
therefore offered as a diagnostic instrument and as triangulation
against the documented record, not as an independently auditable
dataset. Its convergence with the literature mitigates, but does not
eliminate, these effects.

\section{CONCLUSION}

Fifteen years after Dixon\textquotesingle s blog post, the verdict is
sober: for most organizations, the transformation never arrived. The
reported failure rates swing widely, from 60\% to 90\%, across studies
that measure different things and cannot be lined up side by side. What
does not swing is the pattern. Generation after generation of
technology, the rate holds. The Data Lake joins a long list of
transformative tools that delivered less than promised, and the
shortfall was never really about the technology. It was about
prerequisites no one met.

The thesis is simple. You cannot fix organizational dysfunction with a
platform. Fifteen years and several technology generations bear it out.
The Seven Deadly Sins recur across migrations. Governance debt
accumulates whatever the paradigm underneath. The 92\% of executives who
name culture as their main obstacle {[}6{]} will not clear it by
swapping tools.

This matters now because of AI. Organizations building models on
ungoverned data are setting up to repeat the Data Lake story, at higher
cost. The vendor estimates that most AI projects stall before production
{[}62{]} are soft evidence, but they point at the same organizational
deficits that sank the lakes. Until the cultural, skills, and governance
gaps are closed, the money will keep going in ahead of the returns,
whatever the technology is called.

None of this says Data Lakes were a mistake, or that the successors are
empty. The tools work. Organizational readiness is the binding
constraint. Breaking the cycle takes an honest read of organizational
prerequisites before the spend, a standing commitment to governance as
infrastructure rather than overhead, and sober expectations about what
any platform can do without organizational change. Governance is not the
nice-to-have. It is the precondition.

\appendices
\section{Speculative Numeric Instruments}
\label{app:numeric}
The instruments below put governance gravity and reversion into numbers.
We keep them out of the main text on purpose. They rest on thresholds
derived from practitioner experience, not from empirical calibration, and
we offer them as testable hypotheses for future work rather than as
results. A reader should treat every cutoff here as a starting point to
calibrate, not a standard to apply.

\textbf{Governance Parity Gap Metrics.} To make the warehouse-to-lake
parity gap measurable, we propose five candidate metrics: (1) \emph{KPI
divergence rate}, the frequency and magnitude of discrepancies between
the same metric computed from warehouse versus lake, measured monthly,
where sustained divergence above a few percent on critical KPIs usually
blocks migration; (2) \emph{manual reconciliation volume}, the hours per
month spent reconciling lake outputs against the warehouse source of
truth, where high sustained volumes indicate a persistent gap; (3)
\emph{dual-run duration}, the time in parallel operation before
decommission, where very long dual-runs suggest parity was never
reached; (4) \emph{regulatory incident rate}, quality or audit findings
specific to lake-sourced regulatory reports, which often trigger rapid
reversion; and (5) \emph{semantic definition coverage}, the share of
warehouse business terms and metrics formally implemented in the lake
governance layer, where low coverage correlates with high reversion
risk. The point of the set is to make falsifiable claims about
governance gravity possible, not to assert validated cutoffs.

\textbf{A Validation Protocol.} An organization wishing to validate these
metrics in its own setting could establish baselines at project
initiation, track them monthly across at least two reporting cycles,
log reversion events with timestamps (shadow marts created, reports
migrated back, warehouse expansion requests), correlate the metric
trajectories with the timing of those events, and compare predicted with
observed reversion to calibrate the thresholds. A minimal version would
follow a single migration through two quarterly closes, measuring KPI
divergence and reconciliation volume before and after. Cross-organization
validation would need shared metric definitions and anonymized
benchmarking across several comparable implementations.

\textbf{A Reversion Heuristic.} As a single summary signal, we sketch a
decision rule: when several swamp indicators stay in a red zone across
two consecutive quarters, the organization tends to re-institutionalize
warehouse-sourced data marts for mission-critical workloads within a few
quarters. The two-quarter window reflects the budget and planning cycle
in which complaints accumulate and leadership finally acts; the lag to
implementation reflects the time to stand up parallel infrastructure and
revalidate reconciliation. One could fold the same intuition into a
rough severity score that rises with the number of red-zone indicators,
how long they persist, and how far down the degradation pathway the lake
has slid. We state these only as hypotheses: if a future study validated
them, they could power an early-warning signal while remediation is still
feasible.

\end{document}